# On the Complexity of Nonrecursive XQuery and Functional Query Languages on Complex Values

Christoph Koch
DBAI, Institut für Informationssysteme
Technische Universität Wien
A-1040 Vienna, Austria
koch@dbai.tuwien.ac.at

## ABSTRACT

*This paper studies the complexity of evaluating functional query languages for complex values such as monad algebra and the recursion-free fragment of XQuery.*

*We show that monad algebra with equality restricted to atomic values is complete for the class $TA[2^{O(n)}, O(n)]$ of problems solvable in linear exponential time with a linear number of alternations. The monotone fragment of monad algebra with atomic value equality but without negation is complete for nondeterministic exponential time. For monad algebra with deep equality, we establish $TA[2^{O(n)}, O(n)]$ lower and exponential-space upper bounds.*

*Then we study a fragment of XQuery, Core XQuery, that seems to incorporate all the features of a query language on complex values that are traditionally deemed essential. A close connection between monad algebra on lists and Core XQuery (with "child" as the only axis) is exhibited, and it is shown that these languages are expressively equivalent up to representation issues. We show that Core XQuery is just as hard as monad algebra w.r.t. combined complexity, and that it is in $TC_0$ if the query is assumed fixed.*

## 1. INTRODUCTION

Complex values form part of various data models for advanced database applications, such as object-oriented, object-relational, and semistructured data models. A large amount of theoretical work on query languages for complex values has been done (e.g. [26, 29, 1, 24, 21, 39, 20, 25, 6, 2, 36, 43, 10, 32, 41, 11]), and this has laid the foundations for object-oriented query languages as well as SQL 1999 or XQuery.

Earlier complexity studies on query languages for complex values have almost entirely focused on logic- [30] and particularly logic programming-based query languages [41, 11, 9], and fixpoint languages (e.g. [21]). However, the query languages considered by many researchers to be most natural for complex values (such as *complex value algebra without powerset* [1, 3], its syntactic variant *monad algebra* [39, 6], and XQuery) are functional.

**Monad algebra.** Monad algebra was shown expressively equivalent to a number of other important complex-value query languages such as *nested relational algebra* [26] and complex value algebra without powerset in previous research [39]. (Complex value algebra *with powerset* [29, 1, 20] can take hyperexponential runtime. Queries that really need the powerset operator are usually too costly to evaluate.)

Since some of these languages were developed driven by practical requirements rather than from first principles as is the case for monad algebra, it appears that the expressiveness of these languages on complex values is "the right one" to many researchers and plays a role analogous to that of the power of first-order logic (or relational algebra) on the relational model.

One known result [38] is that monad algebra is in $TC_0$ w.r.t. *data complexity* (i.e., if the query is assumed fixed [40]). However, the complexity of monad algebra if the query is assumed variable (*query/combined complexity* [40]). is open. In this paper, we study the complexity of monad algebra under the latter assumption.

**XQuery.** XQuery is destined to become the dominant data-transformation query language for XML data and to take a role analogous to the one occupied by SQL for relational databases.

It is folklore that full XQuery is Turing-complete, but it is also obvious that queries without recursion are guaranteed to terminate in straightforward functional implementations of the XQuery language. Recursion in XQuery is rarely used in practice (see also [45]); recursive XML transformations are usually implemented in XSLT.

In essence, XQuery is a quite natural typed functional programming language for XML; still it is sometimes criticized by the research community as huge and clumsy. In this paper we study a substantial recursion-free fragment of XQuery, which we call Core XQuery[1]. It seems that Core XQuery contains all and only the features one would expect from a functional query language for unranked trees in the spirit of complex-value algebra without powerset.

Little foundational research on XQuery has been done to date. There are only some cautious first attempts at finding clean formalizations of and algebras for the language [23, 12, 44]. Most other recent work has focused on engineering good query processors for XQuery [33, 34, 15, 13, 28].

In this paper, we attempt a first closer look at the complexity of XQuery, or more precisely, of the Core XQuery fragment. We attempt to do this in a principled manner, establishing connections to earlier, well-studied formalisms for functional queries on complex-value databases [36, 39, 6, 43, 19, 32]. Indeed our results on the complexity of monad algebra quite directly yield a characterization of the complexity of Core XQuery.

---
[1] This fragment is not to be confused with the XQuery Core [44], which is a much larger fragment of XQuery that is also harder to study.

|  | with negation | without negation |
|---|---|---|
| deep equality | in EXPSPACE; TA$[2^{O(n)}, O(n)]$-hard | |
| equality on atomic values | TA$[2^{O(n)}, O(n)]$-complete | NEXPTIME-complete |

Table 1: Summary of results on query/combined complexity for monad algebra and Core XQuery.

The technical contributions of this paper are as follows.

- We show that monad algebra on sets, lists and bags is complete for TA$[2^{O(n)}, O(n)]$ w.r.t. combined complexity in the presence of negation and equality on atomic values.

- We show that monad algebra on sets, lists and bags with equality on atomic values but without negation is NEXPTIME-complete.

- For the case of monad algebra with deep equality, we obtain an EXPSPACE upper bound. A TA$[2^{O(n)}, O(n)]$ lower bound follows from the fact that negation is easily definable using deep equality.

- We introduce the Core XQuery language, a simple yet powerful nonrecursive fragment of XQuery.

- We exhibit a close connection between XQuery and monad algebra on lists and show that the Core XQuery queries that use only the child axis for navigation in data trees capture monad algebra on lists up to representation issues.[2]

  The established mappings are efficiently computable. This allows us to make use of our complexity characterization of monad algebra, but it also gives a very concise formal semantics to Core XQuery.

- We show that if equality is restricted to atomic values, Core XQuery is complete for TA$[2^{O(n)}, O(n)]$.

- Core XQuery with deep equality is in EXPSPACE and hard for the class TA$[2^{O(n)}, O(n)]$. Since we can again very directly express negation using deep equality, this result holds even if negation and universal quantification ("every") are ruled out from the language.

- We show that the monotone fragment of Core XQuery – without negation and with equality restricted to atomic values – is NEXPTIME-complete.

- Finally, we show that Core XQuery is in TC$_0$ w.r.t. data complexity.

Table 1 summarizes our complexity results for query and combined complexity.

[2]Monad algebra on nested lists (which are in fact the same as unranked, ordered, *unlabeled* trees) is uncomparable with Core XQuery in the strict sense because we have excluded position arithmetics and attributes from Core XQuery and thus cannot simulate the tuples of monad algebra that are essential to its expressiveness.

To the best of the author's knowledge, this is the first work characterizing the complexity of XQuery. The mappings to and from monad algebra also give an argument that Core XQuery is a well-designed language that offers the "right" degree of expressive power.

**Related work**. It seems that the most relevant work regarding this problem – apart from the characterization of the data complexity of monad algebra in [38] – is on the complexity of nonrecursive logic programming.

For nonrecursive logic programming, a full complexity characterization [9] has been obtained for the most common forms of complex values (that is, values built from sets, lists, bags, tuples, and atomic values.) and various classes of logic programs (with and without negation, range-restriction, and types). It turns out that the complexity of nonrecursive logic programming is robustly (for various kinds of complex values, and with or without range-restriction) NEXPTIME-complete. In the presence of negation (and necessarily range-restriction), nonrecursive logic programming is known to be in the class TA$[2^{O(n)}, O(n)]$ [41] and hard for the class TA$[2^{O(n/\log n)}, O(n/\log n)]$ [42, 9].

A main difference between functional languages such as monad algebra and XQuery and logic programming as studied in [10, 41] is the form of nonmonotonicity employed. In functional languages that have the power to check the equality of complex values, negation is usually a redundant operation. Equality introduces nonmonotonicity into the functional languages, while the seemingly same deep equality is innocuous in logic programming languages. Nonmonotonicity in the functional languages is different and seemingly more powerful than that obtained through negation in nonrecursive normal logic programming. For example, in monad algebra, we can compute two sets of doubly exponential size in two different ways (using a "map" operation that applies a transformation to each element of a set) and then check their equality, while the upper bounds on the complexity of nonrecursive logic programming rely on the fact that unifiers in SLD resolution proofs of nonrecursive logic programs cannot grow beyond singly exponential size.

The work [21, 30] is on more expressive query languages. [30] proves *LDM logic* without powerset complete for the class TA$[2^{O(n)}, O(n)]$. Differently from monad algebra, LDM logic is a logical language with quantification, operates on cyclic data, and cannot express deep equality.

**Structure**. The structure of this paper is as follows. Section 2 discusses some necessary complexity-theoretic background and introduces complex values and monad algebra (on sets). Section 3 studies the complexity of monad algebra on sets, focusing on upper bounds, while Section 4 provides the corresponding lower bounds. Section 5 briefly discusses the complexity of monad algebra on lists and bags. Section 6 defines the Core XQuery fragment and provides efficiently computable mappings between monad algebra on lists and XQuery. Sections 7 and 8 present our results on combined and data complexity of XQuery, respectively.

## 2. PRELIMINARIES

### 2.1 Complexity-Theoretic Background

By $AC_0$ we refer to the class of languages recognizable by LOGSPACE-uniform families of circuits using and- and or-gates of unbounded fan-in of polynomial size and constant depth. By $TC_0$ we refer to the same class except that in addition so-called majority-gates are permitted, which compute "true" iff more than half of their inputs are true. For details on circuit complexity and the notion of uniformity we refer to [18, 27].

We assume deterministic, nondeterministic, and alternating Turing machines known and refer to e.g. [27] for definitions. By $DTIME[t(n)]$ and $NTIME[t(n)]$, we denote the classes of all problems solvable in time $t(n)$ (where $n$ is the size of the input) on deterministic and nondeterministic Turing machines, respectively. By $DSPACE[s(n)]$, we denote the classes of all problems solvable in space $s(n)$ on deterministic Turing machines. By $TA[t(n), a(n)]$, we denote the class of problems solvable in time $t(n)$ using $a(n)$ alternations on alternating Turing machines.

We will use the following abbreviations for complexity classes in this paper:

$$
\begin{aligned}
\text{NETIME} &= \text{NTIME}[2^{O(n)}] \\
\text{NEXPTIME} &= \text{NTIME}[2^{n^{O(1)}}] \\
\text{2ETIME} &= \text{DTIME}[2^{2^{O(n)}}] \\
\text{2EXPTIME} &= \text{DTIME}[2^{2^{n^{O(1)}}}] \\
\text{LOGSPACE} &= \text{DSPACE}[O(\log n)] \\
\text{EXPSPACE} &= \text{DSPACE}[2^{n^{O(1)}}]
\end{aligned}
$$

It is known that $AC_0 \subseteq TC_0 \subseteq \text{LOGSPACE} \subset \text{NEXPTIME} \subseteq TA[2^{n^{O(1)}}, 1] \subseteq TA[2^{n^{O(1)}}, n^{O(1)}] \subseteq TA[2^{n^{O(1)}}, 2^{n^{O(1)}}] = \text{EXPSPACE} \subseteq \text{2EXPTIME}$. Moreover,

$$\text{NETIME} \subseteq TA[2^{O(n)}, O(n)] \subseteq \text{2ETIME} \subset \text{2EXPTIME}$$

(cf. e.g. [27, 7]).

NETIME and 2ETIME are not robust complexity classes – they are not closed under LOGSPACE-reductions, as can be verified using a simple padding argument and the Time Hierarchy theorem [22]. We will consider completeness for those classes as well as of $TA[2^{O(n)}, O(n)]$ under *LOGLIN-reductions*, under which they are known to be closed (cf. e.g. [9]). By a LOGLIN reduction, we denote a LOGSPACE reduction that produces output of linear size. $TA[2^{O(n)}, O(n)]$ is known to be closed under LOGLIN reductions and has important complete problems from logic, such as deciding the Theory of Real Addition [5, 14].

### 2.2 Complex Values and Monad Algebra

We now introduce monad algebra on sets; monad algebra on lists and bags will be briefly sketched in Section 5.

We consider complex values constructed from sets, tuples, and atomic values from a single-sorted domain[3]. Types are terms of the grammar

$$\tau ::= \text{Dom} \mid \{\tau\} \mid \langle A_1 : \tau_1, \ldots, A_k : \tau_k \rangle$$

where $k \geq 0$.

[3] All results in this paper immediately generalize to many-sorted domains.

Consider the query language on complex values consisting of expressions built from the following operations (the types of the operations are provided as well):

1. identity
$$\text{id} : x \mapsto x \qquad \tau \to \tau$$

2. composition[4]
$$f \circ g : x \mapsto g(f(x)) \qquad \frac{f : \tau \to \tau', \; g : \tau' \to \tau''}{f \circ g : \tau \to \tau''}$$

3. constants from $\text{Dom} \cup \{\emptyset, \langle\rangle\}$ ($\langle\rangle$ is the nullary tuple)

4. singleton set construction
$$\text{sng} : x \mapsto \{x\} \qquad \tau \to \{\tau\}$$

5. application of a function to every member of a set
$$\text{map}(f) : X \mapsto \{f(x) \mid x \in X\}$$
$$\frac{f : \tau \to \tau'}{\text{map}(f) : \{\tau\} \to \{\tau'\}}$$

6. flatten: $X \mapsto \bigcup X \qquad \{\{\tau\}\} \to \{\tau\}$

7. pairing
$$\text{pairwith}_{A_1} : \langle A_1 : X_1, A_2 : x_2, \ldots, A_n : x_n \rangle \mapsto$$
$$\{\langle A_1 : x_1, A_2 : x_2, \ldots, A_n : x_n \rangle \mid x_1 \in X_1\}$$

$$\langle A_1 : \{\tau_1\}, A_2 : \tau_2, \ldots, A_n : \tau_n \rangle \to$$
$$\{\langle A_1 : \tau_1, \ldots, A_n : \tau_n \rangle\}$$

(pairwith$_{A_i}$ for $i > 1$ is defined analogously.)

8. tuple formation
$$\langle A_1 : f_1, \ldots, A_n : f_n \rangle :$$
$$x \mapsto \langle A_1 : f_1(x), \ldots, A_n : f_n(x) \rangle$$

$$\frac{f_1 : \tau \to \tau_1, \ldots, f_n : \tau \to \tau_n}{\langle A_1 : f_1, \ldots, A_n : f_n \rangle : \tau \to \langle A_1 : \tau_1, \ldots, A_n : \tau_n \rangle}$$

9. projection
$$\pi_{A_i} : \langle A_1 : x_1, \ldots, A_i : x_i, \ldots, A_n : x_n \rangle \mapsto x_i$$
$$\pi_{A_i} : \langle A_1 : \tau_1, \ldots, A_n : \tau_n \rangle \to \tau_i$$

The language has a nice theoretical foundation from programming language theory, that of structural recursion on sets extended by a small amount of machinery for creating and destroying tuples [39]. Formally, the language above is a Cartesian category with a *strong monad* on it (where "strong" refers to so-called *tensorial strength* introduced by the "pairwith" operation). We call this language *monad algebra* [39], or $\mathcal{M}$ for short.

We will use flatmap($f$) as a shortcut for map($f$) $\circ$ flatten. Observe that projection $\pi$ is applied to tuples rather than to sets of tuples as in relational algebra. For example, the relational algebra expression $\pi_{AB}$ corresponds to the expression map($\langle A : \pi_A, B : \pi_B \rangle$) in $\mathcal{M}$.

[4] Again, our convention throughout the paper is that $(f \circ g)(x) = g(f(x))$, not $f(g(x))$.

EXAMPLE 2.1. The Cartesian product $f \times g$ can be defined as $\langle 1 : f, 2 : g \rangle \circ \text{pairwith}_1 \circ \text{flatmap}(\text{pairwith}_2)$.

Observe the difference from the product of relational algebra. For instance, the query $\text{id} \times \text{id}$ on a set of pairs $S$ computes the set $\{\langle\langle x_1, x_2\rangle, \langle x_3, x_4\rangle\rangle \mid \langle x_1, x_2\rangle, \langle x_3, x_4\rangle \in S\}$ rather than $\{\langle x_1, x_2, x_3, x_4\rangle \mid \langle x_1, x_2\rangle, \langle x_3, x_4\rangle \in S\}$. □

It is customary to define Boolean queries ("predicates") as queries that produce values of type $\{\langle\rangle\}$, i.e., that either return $\{\langle\rangle\}$ ("true") or $\emptyset$ ("false") [39]. Note that the logical conjunction $\gamma \wedge \delta$ of two predicates $\gamma$ and $\delta$ can be computed as $\gamma \times \delta$.

By *positive monad algebra* $\mathcal{M}_\cup$, we denote $\mathcal{M}$ extended by the set union operation $\cup$. This language has a number of nice properties [39, 6], but it is known to be incomplete as a practical query language because it cannot yet express an equality predicate

$$(A_i = A_j) : \langle A_1 : \tau_1, \ldots, A_k : \tau_k\rangle \to \{\langle\rangle\}.$$

However, if we extend $\mathcal{M}_\cup$ by any nonempty subset of the operations equality $(A = B)$, testing set membership $(A \in B)$ or containment $(A \subseteq B)$, selection $\sigma_{A=B}$, set difference "$-$", set intersection $\cap$, or nesting[5], we always get the same expressive power.[6] We will call any one of these extended languages *full monad algebra*.

THEOREM 2.2 ([39]). $\mathcal{M}_\cup[=] \equiv \mathcal{M}_\cup[\sigma] \equiv \mathcal{M}_\cup[-] \equiv \mathcal{M}_\cup[\cap] \equiv \mathcal{M}_\cup[\subseteq] \equiv \mathcal{M}_\cup[\in] \equiv \mathcal{M}_\cup[\text{nest}]$.

Moreover, generalizing selections to test against constants or to support "$\in$", "$\subseteq$", or Boolean combinations of conditions does not increase the expressiveness of full monad algebra [39].

EXAMPLE 2.3. Given a Boolean predicate $\gamma$, selection $\sigma_\gamma$ can be expressed as $\text{flatmap}(\langle 1 : \text{id}, 2 : \text{id} \circ \gamma\rangle \circ \text{pairwith}_2 \circ \text{map}(\pi_1))$.

Predicate $(A \subseteq B)$ can be expressed in $\mathcal{M}_\cup[=]$ as

$$\langle A : \pi_A, A' : \pi_A \cap \pi_B\rangle \circ (A = A')$$

where $f \cap g := (f \times g) \circ \sigma_{1=2} \circ \text{map}(\pi_1)$. A predicate $(f \subseteq g)$ can of course be expressed as $\langle 1 : f, 2 : g\rangle \circ (1 \subseteq 2)$. □

EXAMPLE 2.4. Given a complex value of type

$$\langle R : \{\tau\}, S : \{\tau\}\rangle,$$

difference $R - S$ can be implemented in $\mathcal{M}_\cup[\sigma]$ as

$$\text{pairwith}_R \circ \text{map}\big(\langle R : \pi_R, S_R : \langle R : \pi_R, S : \pi_S\rangle \circ$$
$$\text{pairwith}_S \circ \sigma_{R=S}\rangle\big) \circ \sigma_{S=\emptyset} \circ \text{map}(\pi_R).$$

The idea is to compute, for each element $r$ of $R$, the set $S_R$ of elements in $S$ that are equal to $r$ and then to select those elements $r$ of $R$ for which $S_R$ is empty. □

Theorem 2.2 demonstrates that full monad algebra (w.l.o.g., $\mathcal{M}_\cup[=]$) is a very robust notion. It can serve as an "expressiveness benchmark" for query languages on complex-value databases. Indeed, it has been shown that full monad algebra is a *conservative extension* of relational algebra:[7]

THEOREM 2.5 ([36]). *A mapping from a (flat) relational database to a (flat) relation is expressible in $\mathcal{M}_\cup[=]$ if and only if it is expressible in relational algebra.*

There are a number of alternative ways of stating the query evaluation problem. In this paper, we study the complexity of Boolean queries. For XQuery, we will study the problem of deciding whether the root node (which must be always present in a valid XQuery result) of the resulting XML tree has children.

In the following, we will discuss three kinds of complexity of query evaluation, *data complexity* (where queries are assumed to be fixed and data variable), *query complexity* (where the query is variable and the data is assumed to be fixed), and *combined complexity* (where both data and query are considered variable) [40].

## 3. COMPLEXITY OF MONAD ALGEBRA

We will study the complexity of full monad algebra $\mathcal{M}_\cup[=]$ as well as monotone fragments. It is folklore that by extending $\mathcal{M}_\cup$ by equality on atomic values $=_{atomic}$, we still cannot express nonmonotone operations such as equality of sets or negation. We can safely generalize $=_{atomic}$ to equality of arbitrary complex values that do not include sets, $=_{mon}$, defined inductively as $=_{atomic}$ on atomic values and $v_1 =_{mon} w_1 \wedge \cdots \wedge v_k =_{mon} w_k$ on tuples $\langle v_1, \ldots, v_k\rangle$ and $\langle w_1, \ldots, w_k\rangle$. Of course this generalization does not improve upon the expressiveness of $\mathcal{M}_\cup[=_{atomic}]$.

PROPOSITION 3.1. $\mathcal{M}_\cup[=_{atomic}]$ *captures* $\mathcal{M}_\cup[=_{mon}]$.

**Proof Sketch.** Of course, every $\mathcal{M}_\cup[=_{atomic}]$ query is also a $\mathcal{M}_\cup[=_{mon}]$ query. For the other direction, we can define $=_{mon}$ using $=_{atomic}$ given the type $\tau$ of the values to compare. Viewing each such tuple type as a ranked tree $t$, we simply define $(A =_{mon}^\tau B)$ as the conjunction (implemented as the Cartesian product) of the equality predicates $(A.\pi =_{atomic} B.\pi)$ for each attribute path $\pi$ in $t$ from the root to a leaf. For example, for type

$$\tau = \langle C : \langle D : \text{Dom}, E : \langle F : \text{Dom}, G : \text{Dom}\rangle\rangle, H : \text{Dom}\rangle,$$

$$(A =_{mon}^\tau B) :=$$
$$(A.C.D =_{atomic} B.C.D) \times (A.C.E.F =_{atomic} B.C.E.F) \times$$
$$(A.C.E.G =_{atomic} B.C.E.G) \times (A.H =_{atomic} B.H).$$

(By definition, these types must be constructed from tuples and atomic values.) □

We start our complexity study with data complexity. It is quite easy to conclude from Theorem 2.5 (conservativity over relational algebra) that the data complexity of $\mathcal{M}_\cup[\sigma]$ must be rather low.

PROPOSITION 3.2 (FOLKLORE, [38]). $\mathcal{M}_\cup[=]$ *is in* $TC_0$ *w.r.t. data complexity.*

---

[5] The "nest" operation of complex value algebra without powerset [1] groups tuples by some of their attributes. For example, $\text{nest}_{C=(B)}(R)$ on relation $R(AB)$ computes the value $\{\langle A : x, C : \{\langle B : y\rangle \mid \langle A : x, B : y\rangle \in R\}\rangle \mid (\exists y)\langle A : x, B : y\rangle \in R\}$.
[6] No analogous statement can be made about flat relational algebra.
[7] A generalized version of Theorem 2.5 can be found in [43].

Since the proof in [38] is somewhat involved, we provide an alternative direct proof in the appendix.

We will now show that the query complexity of monad algebra is substantially higher. Actually, it is possible to write queries that compute values of doubly exponential size.

PROPOSITION 3.3. *There is an $\mathcal{M}_\cup$ query $Q$ that computes a value of size $2^{2^{\Omega(|Q|)}}$.*

**Proof.** Consider the query $Q$

$$\phi_{\{0,1\}} \circ \underbrace{(\text{id} \times \text{id}) \circ \cdots \circ (\text{id} \times \text{id})}_{m \text{ times}}$$

where $\phi_{\{0,1\}} = (0 \circ \text{sng}) \cup (1 \circ \text{sng})$ computes the set $\{0,1\}$ and $m$ is linear in $|Q|$. Query $Q$ computes the set of all nested pairs (=binary trees) of depth $m$ with labels from $\{0,1\}$ at the leaves. There are $2^{2^m}$ such nested pairs. □

For the converse,

PROPOSITION 3.4. *The values computable by $\mathcal{M}_\cup[=]$ queries are of size $2^{2^{O(n)}}$, where $n$ is the size of the input (i.e., database and query).*

**Proof.** Let $C_f(n)$, for each $\mathcal{M}_\cup[=]$ expression $f$, be defined as follows: For constants, it is $O(1)$; for the operation id, it is $|n|$; for sng, it is $|n|+O(1)$; for flatten, $\sigma$, and $\pi$, it is $|n|$, for pair construction $\langle 1 : f, 2 : g \rangle$, it is $C_f(n)+C_g(n)+O(1)$; for union $f \cup g$, it is $C_f(n)+C_g(n)$; for pairwith, it is $n^2+O(1)$; and finally, for $f \circ g$, it is $C_g(C_f(n))$.

It is easy to see that $C_f$ provides us with an upper bound on the size of the value obtained by applying $\mathcal{M}_\cup[=]$ expression $f$ on a value of size $n$.

For $n > 1$, pairwith is the locally costliest operation, so let us assume that $Q$ consists of the composition of $|Q|$ operations with this cost as an upper bound. In particular, this will provide an overestimation of the size of the computed value because for $n > 1$, $C_f(n) + C_g(n) + O(1) < C_{\underbrace{\text{pairwith} \circ \cdots \circ \text{pairwith}}_{|f|+|g| \text{ times}}}(n) = (\cdots((n^2+O(1))^2+O(1))\cdots)^2 + O(1)$. Now,

$$\begin{aligned}
|[\![Q]\!](D)| &\leq C_Q(|D|) \\
&\leq \overbrace{(\cdots(((|D|^2+O(1))^2+O(1))\cdots)^2}^{|Q| \text{ times}} + O(1) \\
&\leq \overbrace{(\cdots((((|D|+O(|Q|))^2)^2)\cdots)^2}^{|Q| \text{ times}} \\
&\leq 2^{2^{O(|D|+|Q|)}}
\end{aligned}$$

□

COROLLARY 3.5. *$\mathcal{M}_\cup[=]$ is in 2ETIME w.r.t. combined complexity.*

This is easy to see because given an input value of size $2^{2^{O(n)}}$, each operation of $\mathcal{M}_\cup[=]$ can be evaluated on the input in time $2^{2^{O(n)}}$ on a random access machine. There are $|Q| \leq n$ operations, and $|Q| \cdot 2^{2^{O(n)}} = 2^{2^{O(n)}}$.

Since monad algebra has the power to construct arbitrary values from scratch, we will use the following proposition and will subsequently focus on query complexity. As all complexity classes we will consider for query and combined complexity throughout this paper will be closed under LOGLIN-reductions, combined complexity is no harder than query complexity.

PROPOSITION 3.6. *There is a LOGLIN reduction that, given a complex value $v$, computes an $\mathcal{M}_\cup$ expression that evaluates to $v$ on an arbitrary (e.g. empty) database.*

The main upper bound results of this paper follow next.

THEOREM 3.7. *$\mathcal{M}_\cup[=_{atomic}]$ is in NEXPTIME w.r.t. query complexity.*

**Proof Sketch.** Without loss of generality, we may assume that all operations of the given monad algebra query are unary. This requires only a slight change of notation when we use the union operation $\cup$: Rather than writing $f \cup g$, we write $\langle A : f, B : g \rangle \circ \cup$.

The proof is by a LOGSPACE-reduction to the *success problem* of nonrecursive logic programming with function symbols (but without sets), i.e. the problem of deciding whether a distinguished boolean predicate evaluates to true. This problem is known to be NEXPTIME-complete [10].

We now come to address the observation made in the introduction that while monad algebra queries may compute complex values of doubly exponential size (Proposition 3.3), resolution proofs for nonrecursive logic programs are always of only singly exponential size [10]. We show that every $\mathcal{M}_\cup[=_{atomic}]$ query can be reduced to a nonrecursive logic program with a single binary function symbol $f$. For convenience, we write terms built using $f$ in a contrived list representation (paths); for example, the term $f(f(x,y),f(z,f(u,v)))$ will be written as $(x.y).z.u.v$. Left $f$-term children are considered Skolem functions generating new path labels. For example, $(x.y).z.u.v$ is understood as a path $w.z.u.v$ where $w$ is a label generated from and identified by $x.y$.

We view every complex value as a deterministic tree, i.e., a tree in which each node $v$ is uniquely identified by the path of labels from the root to $v$. We are able to uniquely assign such labels – even the elements of an index set to the elements of a set value, as we are considering query complexity and construct every set value from scratch (see Proposition 3.6). Such a deterministic tree is of course fully described by the set of root-to-leaf paths occurring in it.

We can now give an alternative semantics of $\mathcal{M}_\cup[=_{atomic}]$ in terms of deterministic trees, that is, each query maps a deterministic tree given as a set of paths to a deterministic tree given as a set of paths, with

$$\begin{aligned}
[\![\text{id}]\!](V) &:= V \\
[\![c]\!](V) &:= \{c\} \\
[\![\pi_A]\!](V) &:= \{v \mid A.v \in V\} \\
[\![\text{sng}]\!](V) &:= \{s.v \mid v \in V\} \\
[\![f \circ g]\!](V) &:= [\![g]\!]([\![f]\!](V)) \\
[\![\text{flatten}]\!](V) &:= \{(i.j).v \mid i.j.v \in V\} \\
[\![A =_{atomic} B]\!](V) &:= \{\langle\rangle \mid A.v, B.v \in V\}
\end{aligned}$$

$[\![\pi_A \cup \pi_B]\!](V) :=$
$\quad \{(1.i).v \mid A.i.v \in V\} \cup \{(2.i).v \mid B.i.v \in V\}$

$[\![\langle A_1 : f_1, \ldots, A_k : f_k \rangle]\!](V) :=$
$\quad \{A_1.v_1, \ldots, A_k.v_k \mid v_1 \in [\![f_1]\!](V) \wedge \cdots \wedge v_k \in [\![f_k]\!](V)\}$

$[\![\mathrm{map}(f)]\!](V) :=$
$\qquad \{i.w \mid \exists u : i.u \in V \wedge w \in [\![f]\!](\{v \mid i.v \in V\})\}$

$[\![\mathrm{pairwith}_{A_j}]\!](V) := \{i.A_j.v \mid A_j.i.v \in V\} \cup$
$\qquad \{i.A_k.w \mid A_j.i.v, A_k.w \in V \wedge j \neq k\}$

Here, $V$ always denotes a set of paths, $u, v, w, v_1, \ldots, v_k$ denote paths, and $i, j$ denotes indexes of set members. The symbol $\langle\rangle$ in the definition for the equality predicate is to be understood as a constant and a path of length one. Observe how the flatten operation merges two set member indexes $i$ and $j$ into one exploiting our *binary* function symbol for encoding paths.

An example demonstrating the construction of

$\{0,1\} \circ (\mathrm{id} \times \mathrm{id}) = \langle 1 : 0 \circ \mathrm{sng}, 2 : 1 \circ \mathrm{sng}\rangle \circ \cup \circ$
$\langle A : \mathrm{id}, B : \mathrm{id}\rangle \circ \mathrm{pairwith}_A \circ \mathrm{map}(\mathrm{pairwith}_B) \circ \mathrm{flatten}$

is shown in Figure 1. This query evaluates to a deterministic tree that can be uniquely specified by its set of root-to-leaf paths $\{((1.s).1.s).A.0, ((1.s).1.s).B.0, \ldots, ((2.s).2.s).B.1\}$.

The reduction of monad algebra queries to nonrecursive logic programming is now technical but not difficult.

Our predicates are binary and of the form $p(X, v)$, where $X$ is a path prefix identifying a node $w$ of the deterministic tree representation of our complex value, and $v$ denotes one of the paths to leaves emanating from $w$, which taken together fully specify the complex value below node $w$.

- We translate an expression $\mathrm{map}(f)$ on path $X$ represented by predicate $[\![Q]\!]$ into

  $[\![Q; \mathrm{start\_map}]\!](X.i, v) \leftarrow [\![Q]\!](X, i.v).$
  $[\![Q; \mathrm{map}(f)]\!](X, i.v) \leftarrow [\![Q; \mathrm{start\_map}; f]\!](X.i, v).$

  plus the translation of $f$ mapping from predicate $[\![Q; \mathrm{start\_map}]\!]$ to $[\![Q; \mathrm{start\_map}; f]\!]$.

  That is, on a value identified by path prefix $X$, we move down to the set member children of $X$, the $X.i$. Then we apply $f$ on the values $X.i$, and finally, we return to $X$.

- We translate an expression $\langle A_1 : f_1, \ldots, A_k : f_k\rangle$ on path $X$ represented by predicate $[\![Q]\!]$ into

  $[\![Q; \langle A_1 : f_1, \ldots, A_k : f_k\rangle]\!](X, A_1.v) \leftarrow [\![Q; f_1]\!](X, v)$
  $\qquad \vdots$
  $[\![Q; \langle A_1 : f_1, \ldots, A_k : f_k\rangle]\!](X, A_k.v) \leftarrow [\![Q; f_k]\!](X, v)$

  plus, for each $1 \leq i \leq k$, the translation of $f_i$ mapping from predicate $[\![Q]\!]$ to predicate $[\![Q; f_i]\!]$.

- Compositions $f \circ g$, are read as $f; g$ and $f$ and $g$ are translated separately. The result predicate of $f$ is used as the input predicate of $g$.

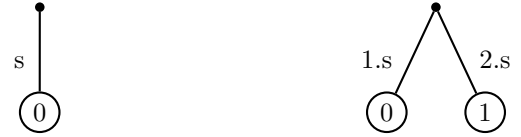

(a): $0 \circ \mathrm{sng}$      (b): $(0 \circ \mathrm{sng}) \cup (1 \circ \mathrm{sng})$

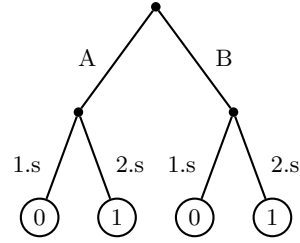

(c): $((0 \circ \mathrm{sng}) \cup (1 \circ \mathrm{sng})) \circ \langle A : \mathrm{id}, B : \mathrm{id}\rangle$

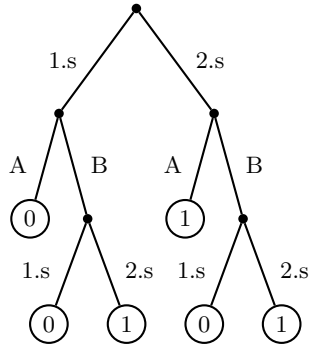

(d): $((0 \circ \mathrm{sng}) \cup (1 \circ \mathrm{sng})) \circ \langle A : \mathrm{id}, B : \mathrm{id}\rangle \circ \mathrm{pairwith}_A$

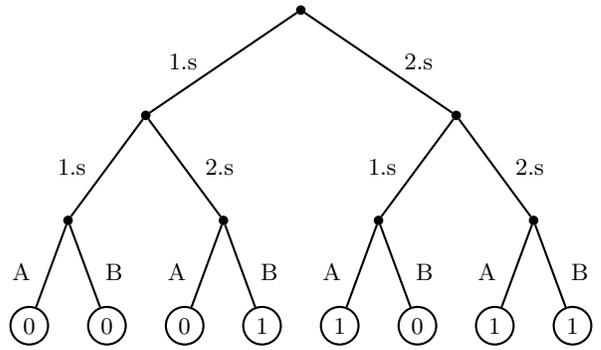

(e): $((0 \circ \mathrm{sng}) \cup (1 \circ \mathrm{sng})) \circ \langle A : \mathrm{id}, B : \mathrm{id}\rangle \circ \mathrm{pairwith}_A \circ \mathrm{map}(\mathrm{pairwith}_B)$

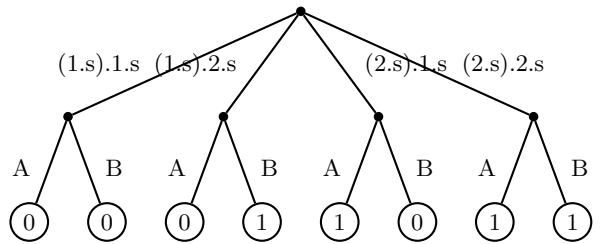

(f): $((0 \circ \mathrm{sng}) \cup (1 \circ \mathrm{sng})) \circ \langle A : \mathrm{id}, B : \mathrm{id}\rangle \circ \mathrm{pairwith}_A \circ \mathrm{map}(\mathrm{pairwith}_B) \circ \mathrm{flatten}$

**Figure 1: Construction of tree $\{0,1\} \circ (\mathbf{id} \times \mathbf{id})$.**

- The remaining operations are translated as follows.

$$[\![Q;c]\!](X,c) \leftarrow [\![Q]\!](X,v).$$

$$[\![Q;\text{pairwith}_B]\!](X,i.B.v) \leftarrow [\![Q]\!](X,B.i.v).$$
$$[\![Q;\text{pairwith}_B]\!](X,i.A.v) \leftarrow [\![Q]\!](X,A.v),$$
$$[\![Q]\!](X,B.i.w).$$

$$[\![Q;\text{flatten}]\!](X,(i.j).v) \leftarrow [\![Q]\!](X,i.j.v).$$

$$[\![Q;(A=_{atomic}B)]\!](X,s.\langle\rangle) \leftarrow [\![Q]\!](X,A.v),$$
$$[\![Q]\!](X,B.v).$$

$$[\![Q;\pi_A \cup \pi_B]\!](X,(1.i).v) \leftarrow [\![Q]\!](X,A.i.v)$$
$$[\![Q;\pi_A \cup \pi_B]\!](X,(2.i).v) \leftarrow [\![Q]\!](X,B.i.v)$$

$$[\![Q;\pi_{A_i}]\!](X,v) \leftarrow [\![Q]\!](X,A_i.v)$$

$$[\![Q;\text{sng}]\!](X,s.v) \leftarrow [\![Q]\!](X,v).$$

By Proposition 3.6, we may assume that our query ignores the input data; so we assume a predicate $[\![\epsilon]\!]$ and a fact $[\![\epsilon]\!](\epsilon,\text{dummy}) \leftarrow .$ as part of our logic program.

It is not hard to verify that this translation of a query $Q$ in $\mathcal{M}_\cup[=_{atomic}]$ into a nonrecursive logic program can be effected in LOGSPACE and that indeed the goal $[\![Q]\!](\epsilon,i.\langle\rangle)$ is true iff $Q$ evaluates to true. □

We consider two examples to illustrate the construction of the logic programs. The save some space, however, we use short predicate names $p_i$.

EXAMPLE 3.8. The logic program for the query $\langle 1:0 \circ \text{sng}, 2:1 \circ \text{sng}\rangle \circ \cup$ is

$$\begin{array}{rcll}
[\![\epsilon]\!](\epsilon,\text{dummy}) & \leftarrow & . & \\
p_1(X,0) & \leftarrow & [\![\epsilon]\!](X,v). & \text{\# constant 0} \\
p_2(X,s.v) & \leftarrow & p_1(X,v). & \text{\# sng} \\
p_3(X,1) & \leftarrow & [\![\epsilon]\!](X,v). & \text{\# constant 1} \\
p_4(X,s.v) & \leftarrow & p_3(X,v). & \text{\# sng} \\
p_5(X,1.v) & \leftarrow & p_2(X,v). & \text{\# create\_tuple} \\
p_5(X,2.v) & \leftarrow & p_4(X,v). & \text{\# create\_tuple} \\
p_6(X,(1.i).v) & \leftarrow & p_5(X,1.i.v). & \text{\# union} \\
p_6(X,(2.i).v) & \leftarrow & p_5(X,2.i.v). & \text{\# union}
\end{array}$$

The goal predicate $p_6$ computes the sets of paths of the deterministic tree representation of the result value, that is, $\{\pi \mid p_6(\epsilon,\pi) \text{ is true}\} = \{(1.s).0,(2.s).1\}$ (see Figure 1 (b)). □

EXAMPLE 3.9. On values of type $\{\langle A:\text{Dom}, B:\text{Dom}\rangle\}$ represented by predicate $p_{input}$, the query

$$\text{map}(\langle C:\pi_A, D:\pi_B \circ \text{sng}\rangle)$$

is encoded as the logic program

$$\begin{array}{rcll}
p_1(X.i,v) & \leftarrow & p_{input}(X,i.v). & \text{\# begin\_map} \\
p_2(X,v) & \leftarrow & p_1(X,A.v). & \text{\# } \pi_A \\
p_3(X,v) & \leftarrow & p_1(X,B.v). & \text{\# } \pi_B \\
p_4(X,s.v) & \leftarrow & p_3(X,v). & \text{\# sng} \\
p_5(X,C.v) & \leftarrow & p_2(X,v). & \text{\# create\_tuple} \\
p_5(X,D.v) & \leftarrow & p_4(X,v). & \text{\# create\_tuple} \\
p_6(X,i.v) & \leftarrow & p_5(X.i,v). & \text{\# end\_map}
\end{array}$$

with goal $p_6$. □

The reduction to nonrecursive logic programming of the proof of Theorem 3.7 can be rather easily extended to a reduction from $\mathcal{M}_\cup[=_{atomic},not]$ to nonrecursive normal logic programming (that is, with negation). All we need to do is encode the operation "not" as

$$[\![Q;\text{not}]\!](X,s.\langle\rangle) \leftarrow \text{set}[\![Q]\!](X),$$
$$\text{not nonempty}[\![Q]\!](X).$$
$$\text{nonempty}[\![Q]\!](X) \leftarrow [\![Q]\!](X,v).$$

where the "set$[\![Q]\!]$" predicates are defined alongside the $[\![Q]\!]$ predicates such that set$[\![Q]\!](X)$ is true iff $X$ is the path prefix of a set, empty or not. This reduction is not in LOGLIN because of the size of the predicates generated. Even if we replace the predicate names by shorter ones of the form $p_i$ (where $i$ is an integer), they are of size $\log n$ each (where $n$ is the size of the input query in monad algebra) and the overall size of the logic program is $O(n \cdot \log n)$. (There are linearly many rules.) But since we can compose this preparation with an ATM run and nonrecursive range-restricted normal logic programming is known to be in TA$[2^{O(n)},O(n)]$ [41], this shows that

COROLLARY 3.10. $\mathcal{M}_\cup[=_{atomic},not]$ is in $TA[2^{O(n\cdot\log n)},O(n\cdot\log n)]$ w.r.t. query complexity.

We can improve this to

THEOREM 3.11. $\mathcal{M}_\cup[=_{atomic},not]$ is in $TA[2^{O(n)},O(n)]$ w.r.t. query complexity.

**Proof Sketch**. The proof is direct, using alternating Turing machines, but again incorporates the deterministic tree technique and the idea of evaluating a logic program, now with negation. We will sketch a fixed alternating Turing machine $M$ that recognizes the $\mathcal{M}_\cup[=]$ queries that evaluate to true. Consider the proof of NEXPTIME-membership of nonrecursive logic programming without negation of [10]. It is by an argument that SLD resolution for such a program is in NEXPTIME because we can start from the goal and then always guess and verify unifiers until we have a proof that the goal is true. Unifiers are of singly exponential size; this is particularly easy to see for the special programs produced in the proof of Theorem 3.7, because there all predicates are over paths, and each path is of size $O(n \cdot 2^{O(n)}) = 2^{O(n)}$, where $n$ is the size of the input query. (There are $O(n)$ steps in the paths and each step is a value of size $2^{O(n)}$ - no greater tuples can be computed by a $\mathcal{M}_\cup[=]$ query of size $n$, and thus by our logic programs.)

Our ATM $M$ basically follows such a resolution strategy to prove that the query evaluates to true. We first compute the logic program of Theorem 3.7 (and its extension to support negation described above) and write it to our worktape. Then we start proving the goal $[\![Q]\!](\epsilon,i.\langle\rangle)$. Inductively, to prove a goal, for a given unifier, we guess a rule, adapt our unifier to the body atoms of the rule (both using existential configurations of the ATM), and then branch out using universal ATM computation to check the body atoms of the rule in parallel. Whenever we encounter a negated atom in a rule body, we employ universal computation to verify that this atom cannot become true. Constructed appropriately, $M$ of course accepts if and only if the goal $[\![Q]\!](\epsilon,i.\langle\rangle)$ is true, and thus iff our $\mathcal{M}_\cup[=]$ query evaluates to true.

Let us study the depth of the computation trees of $M$. The depth of the proof tree of the logic program is only linear in the size of the query. The paths in the computation tree of $M$ are of length $2^{O(n)}$ because all we need to do is choose rules and unify very special terms (our deterministic tree paths) of size $2^{O(n)}$. (One can verify by inspection of the construction of the logic program that this is feasible in linear time in the size of the paths.) The number of alternations used is bounded by the number of predicates in the program (There are $O(n)$ many because there are linearly many rules.) plus the number of negation symbols in the program, which is again $O(n)$. Thus, $M$ is an ATM that runs in time $2^{O(n)}$ with $O(n)$ alternations, and our result is shown. □

REMARK 3.12. The previous proof amounts in no way to a claim that we can close any gap we want – here, the gap between $TA[2^{O(n)}, O(n)]$ and $TA[2^{O(n \cdot \log n)}, O(n \cdot \log n)]$ – by just claiming that there is an appropriate Turing machine that performs our reduction *and* then solves the problem in the desired complexity class. But our proof shows that the predicate names introduced by our reduction from monad algebra to logic programming occupy space while not contributing to the power of the logic programs. This suggests that nonrecursive logic programming "wastes" some succinctness. This is also supported by the fact that, because of the space blow-up caused by the predicates, there is a gap between the currently best known upper bound on the complexity of normal logic programming of $TA[2^{O(n)}, O(n)]$ [41] and the best lower bound of $TA[2^{O(n/\log n)}, O(n/\log n)]$ [42, 9]. □

THEOREM 3.13. $\mathcal{M}_\cup[=]$ *is in EXPSPACE w.r.t. combined complexity.*

**Proof Sketch**. We lack the space to prove this, but a brief argument can be given. Since all values computable in $\mathcal{M}_\cup[=]$ are of at most doubly exponential size (see Proposition 3.4), we can represent an index of a set member (or even a path in the deterministic tree representation of complex values of the proof of Theorem 3.7) in a "register" of singly exponential size. We only use polynomially many (in the size of the query) such registers to evaluate the query using a strategy of recomputation of values on demand.

For example, an operation $\text{map}(f)$ on a subvalue identified by path prefix $\pi$ can be executed by computing each of the path prefixes $\pi.i$, where $i$ identifies an element of the set $\pi$, (we can do this anytime we want because we know the query) and applying $f$ to each of the $\pi.i$.

Deep equality of values identified by path prefixes $\pi_1$ and $\pi_2$ can be checked by verifying for each value identified by path prefix $\pi_1.i$ whether there is an *equal* value identified by some path prefix $\pi_2.j$, and vice-versa. Equality here in general is again deep, so we must employ this procedure recursively, but only up to the at most linear depth of the values; thus we only need linearly many exp-sized registers for checking deep equality. □

## 4. LOWER BOUNDS

In this section we establish lower bounds matching the upper bounds of Theorems 3.7 and 3.11.

THEOREM 4.1. $\mathcal{M}_\cup[=_{atomic}]$ *is NEXPTIME-hard w.r.t. query complexity.*

**Proof Sketch**. The proof is by a LOGSPACE-reduction from NEXPTIME Turing machine acceptance. Let $M = (Q^M, q_0^M, \delta^M, F^M)$ be a nondeterministic Turing machine (NTM) that runs in time $2^{n^{O(1)}}$ on inputs of size $n$. We simulate the computation of $M$ in $\mathcal{M}_\cup[=_{atomic}]$. Each run of $M$ is a sequence of configurations of length $2^{K(n)}$, for a suitable $k$ and $K(n) = n^k$. (We may assume w.l.o.g. that terminating computation paths of $M$ remain in a final state until time $2^{K(n)}$ by appropriate design of $M$.) Each configuration of $M$ consists of a read/write tape, a current state, and a position marker on the tape. Of course every $2^{K(n)}$ time NTM computation uses tape space bounded by $2^{K(n)}$.

There are two main difficulties that we face in this reduction: We have to (i) deal with Turing machine tapes and configurations of exponential size and have to (ii) model the accepting computations of $M$ of exponential length succinctly – the $\mathcal{M}_\cup[=_{atomic}]$ query that must achieve this has to be computable in LOGSPACE and thus must be of polynomial size.

**Modeling configurations**.

- Each tape of a configuration is modeled as a tuple of arity $2^{K(n)}$ (or more precisely, nested pairs of nesting depth $K(n)$) of tape symbols.

  Let $\Sigma = \{s_1, \ldots, s_c\}$ be the (fixed) tape alphabet of $M$. Rather than representing the current position of the read/write head on the tape separately from the tape, we will assume a *valid tape* over extended tape alphabet $\Sigma' = \Sigma \cup \{\triangleright s \triangleleft \mid s \in \Sigma\}$ to contain a single symbol $\triangleright s \triangleleft$ (with $s \in \Sigma$) on the tape that indicates that this tape position stores symbol $s$ and is the current position of the read/write head.

  We can compute the set of all $(2 \cdot c)^{2^{K(n)}}$ such tapes in $\mathcal{M}_\cup$ as
  $$\textit{Tapes} := \phi_{\Sigma'} \circ \underbrace{(\text{id} \times \text{id}) \circ \cdots \circ (\text{id} \times \text{id})}_{K(n) \text{ times}}$$
  where $\phi_{\Sigma'}$ is an appropriate $\mathcal{M}_\cup$ expression that computes $\Sigma'$.[8]

  As a result of this construction, some elements of set *Tapes* do not correspond to valid Turing tapes because they contain either zero or more than two markers indicating the current position of the read/write head on the tape. We will deal with this later.

- A superset of all possible configurations is
  $$\textit{Configs} := (\textit{Tapes} \times Q^M) \circ \text{map}(\langle t : \pi_1, q : \pi_2 \rangle).$$

- The start configuration, consisting of the input tape, the start state, and the position marker at position 0 of the tape is obtained as follows.

  We compute the start tape as the input $x$, with $|x| = n$, padded with $(2^{K(n)} - n)$ #-symbols (denoting unused tape space) and with the first position marked, but in our nested pairs representation.

  Let query $\phi_x$ define the nested pair of depth $\lceil \log_2 n \rceil$ representing $x$ padded by $(2^{\lceil \log_2 n \rceil} - n)$ #-symbols, and with the first position marked. (This is easy to

---

[8]I.e., $\phi_{\Sigma'} := s_1 \circ \text{sng} \cup \cdots \cup s_c \circ \text{sng} \cup \triangleright s_1 \triangleleft \circ \text{sng} \cup \cdots \cup \triangleright s_c \triangleleft \circ \text{sng}$.

compute in LOGSPACE.) For example, for input $x = 01101$, the value computed[9] is

$$\langle\langle\langle\triangleright 0\triangleleft, 1\rangle, \langle 1, 0\rangle\rangle, \langle\langle 1, \#\rangle, \langle\#, \#\rangle\rangle\rangle.$$

The start tape is

$$\phi_{start} := \langle 1 : \phi_x, 2 : \phi_{empty}\rangle \circ \underbrace{\phi_{pad} \circ \cdots \circ \phi_{pad}}_{K(n) - \lceil \log_2 n \rceil - 1 \text{ times}}$$

with

$$\phi_{pad} = \langle 1 : \text{id}, 2 : \langle 1 : \pi_2, 2 : \pi_2\rangle\rangle,$$

$$\phi_{empty} := \# \circ \underbrace{\langle \text{id}, \text{id}\rangle \circ \langle \text{id}, \text{id}\rangle \circ \cdots \circ \langle \text{id}, \text{id}\rangle}_{\lceil \log_2 n \rceil \text{ times}}.$$

This takes the value computed by $\phi_x$ – which contains the input and some padding up to $2^{\lceil \log_2 n \rceil}$ symbols, pairs it with a sequence of #-symbols of the same length (computed by $\phi_{empty}$), and then iteratively doubles the length of the tape by appending two copies of the second half of the already computed tape (because the second half consists exclusively of #-symbols). By this trick, there is a *fixed* expression $\phi_{pad}$ that we can compose our query with to double the length of the value produced.

The start configuration is

$$C_{start} := \langle t : \phi_{start}, q : q_0^M\rangle.$$

Observe that $C_{start}$ is a valid configuration with precisely one tape head position marker.

- The accepting configurations are those configurations in which the state is an element of the set $F^M = \{f_1, \ldots, f_{|F^M|}\}$ of accepting states of $M$:

  $AcceptingConfigs :=$
  $Configs \circ (\sigma_{q=_{atomic} f_1} \cup \cdots \cup \sigma_{q=_{atomic} f_{|F^M|}}).$

- In the following, we will test equality of nested pairs (tape segments) and configurations of exponential size. We can define an equality test $=_{mon}$ of linear size on tapes and tape segments using only $=_{atomic}$ inductively as follows. On values of type Dom, $=_{mon}$ is $=_{atomic}$. Otherwise, on pairs $\langle 1 : \tau_1, 2 : \tau_2\rangle$,

  $(A =_{mon} B) := ((\pi_A \circ \phi) \times (\pi_B \circ \phi)) \circ$
  $\sigma_{1.T=_{atomic} 2.T} \circ \sigma_{1.V=_{mon} 2.V} \circ (\text{id} \times \text{id}) \circ$
  $\sigma_{1.1.T=_{atomic}} \text{"1"} \circ \sigma_{2.1.T=_{atomic}} \text{"2"} \circ \text{map}(\langle\rangle)$

  where $\phi := (\langle T : 1, V : \pi_1\rangle \circ \text{sng} \cup \langle T : 2, V : \pi_2\rangle \circ \text{sng})$. For configurations $C, C'$,

  $(C =_{mon} C') \Leftrightarrow (C.t =_{mon} C'.t \wedge C.q =_{atomic} C'.q).$

- Next we define an $\mathcal{M}_\cup[=_{atomic}]$ expression $\phi_{succ}$ that computes the pairs of configurations $\langle C, C'\rangle$ such that $C'$ is a possible successor of $C$, i.e., computable using the transition relation $\delta^M$ of $M$ in one step.

---

[9]It may be advisable to have a special symbol indicating the left end of the tape on its leftmost position to help the machine avoid running out of bounds. We assume such a symbol part of the input, rather than of our construction.

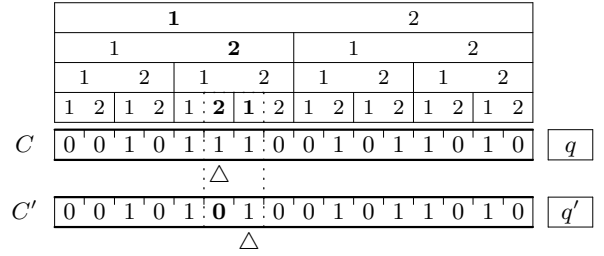

$(q', 0, +1) \in \delta^M(q, 1)$

**Figure 2: Zooming into the tapes to find a valid tape change resulting from a computation step of $M$.**

Here the exponential size of the configurations is a problem; $\phi_{succ}$ has to be chosen carefully in order not to be of exponential size. We achieve this as follows. We start with the Cartesian product of *Configs* – all pairs of configurations, even many that are invalid because they have zero or more than two head markers. For each pair, we make working copies $w, w'$ of the tapes. We achieve this by $\mathcal{M}_\cup$ expression

$$\phi_{prepare-succ} := \textit{Configs} \circ (\text{id} \times \text{id}) \circ$$
$$\text{map}(\langle s : \text{id}, w : \pi_{C.t}, w' : \pi_{C'.t}\rangle).$$

For $w'$ to be a possible successor of $w$, the two tapes may differ at at most two consecutive tape positions (if the tape head moved, otherwise they may only differ at at most one position), and these positions must contain the read/write head position marker. We synchronously "zoom into" the working copies to find these two positions using the following three rules:

1. If $w.2 = w'.2$ (i.e., the second halves of the tapes are equal), replace $w$ by $w.1$ and $w'$ by $w'.1$.

2. If $w.1 = w'.1$ (i.e., the first halves of the tapes are equal), replace $w$ by $w.2$ and $w'$ by $w'.2$.

3. If $w.1.1 = w'.1.1$ and $w.2.2 = w'.2.2$ (i.e., the first and last quarters of the tapes are equal) replace $w$ by newly constructed pair $\langle 1 : w.1.2, 2 : w.2.1\rangle$ and $w'$ by $\langle 1 : w'.1.2, 2 : w'.2.1\rangle$ (that is, by the second and third quarters).

All three cases may apply at the same time because the tapes of two valid configurations $C, C'$, where $C'$ is a successor of $C$, can be equal. An example of iterative zooming is shown in Figure 2. There, we look at a nested pair term of depth four (covering a tape of length 16) and the tape change occurs at positions 6 and 7. We first zoom into the left (using Rule 1) and from there into the right half (using Rule 2). Now both halves differ, but the first and fourth quarter do not, so we can use Rule 3 to zoom down to the differing tape positions 6 and 7. In general, we obtain a tape sequence of length two by zooming into a tape (which is of length $2^{K(n)}$) $K(n) - 1$ times.

In our encoding in $\mathcal{M}_\cup[=_{atomic}]$, we will compute the union of all triples $(s, w, w')$ such that $s$ is a pair of configurations with tapes $t = uwv$ and $t = uw'v$ and $w$ and $w'$ are of length 2 (i.e., $w$ and $w'$ are – if any

– the only corresponding sequences in $t, t'$ that differ). Now we have to make sure that $w$ and $w'$ contain the position marker.

This can be expressed in $\mathcal{M}_\cup[=_{atomic}]$ as follows:

$$\phi_{witness-succ} := \phi_{prepare-succ} \circ \underbrace{\phi_{zoom-in} \circ \cdots \circ \phi_{zoom-in}}_{K(n)-1 \text{ times}} \circ \phi_{marker}$$

where[10]

$$\begin{aligned}
\phi_{zoom-in} &:= (\sigma_{12\triangleright 34\triangleleft} \circ \pi_{12\triangleright 34\triangleleft} \cup \\
&\quad \sigma_{\triangleright 12\triangleleft 34} \circ \pi_{\triangleright 12\triangleleft 34} \cup \sigma_{1\triangleright 23\triangleleft 4} \circ \pi_{1\triangleright 23\triangleleft 4}) \\
\sigma_{12\triangleright 34\triangleleft} &:= \sigma_{w.1=_{mon}w'.1} \\
\pi_{12\triangleright 34\triangleleft} &:= \text{map}(\langle s: \pi_s, w: \pi_{w.2}, w': \pi_{w'.2}\rangle) \\
\sigma_{\triangleright 12\triangleleft 34} &:= \sigma_{w.2=_{mon}w'.2} \\
\pi_{\triangleright 12\triangleleft 34} &:= \text{map}(\langle s: \pi_s, w: \pi_{w.1}, w': \pi_{w'.1}\rangle) \\
\sigma_{1\triangleright 23\triangleleft 4} &:= \sigma_{w.1.1=_{mon}w'.1.1} \circ \sigma_{w.2.2=_{mon}w'.2.2} \\
\pi_{1\triangleright 23\triangleleft 4} &:= \text{map}(\langle s: \pi_s, \\
&\qquad w: \pi_w \circ \langle 1: \pi_{1.2}, 2: \pi_{2.1}\rangle, \\
&\qquad w': \pi_{w'} \circ \langle 1: \pi_{1.2}, 2: \pi_{2.1}\rangle\rangle)
\end{aligned}$$

and $\phi_{marker}$ selects those tuples for which $w, w'$ are two tapes of length two that contain the read/write head marker:

$$\begin{aligned}
\phi_{marker} &:= (\sigma_{w.1=_{atomic}\triangleright s_1 \triangleleft} \cup \cdots \cup \sigma_{w.1=_{atomic}\triangleright s_c \triangleleft} \cup \\
&\quad \sigma_{w.2=_{atomic}\triangleright s_1 \triangleleft} \cup \cdots \cup \sigma_{w.2=_{atomic}\triangleright s_c \triangleleft}).
\end{aligned}$$

Now, for each $\langle s: X, w: Y, w': Z\rangle \in [\![\phi_{witness-succ}]\!]$, either $C = C'$ or $Y, Z$ are precisely the at most two adjacent positions of the tapes of $C$ and $C'$ that can differ if $C'$ is to be a successor of $C$. We can easily encode the valid successors with respect to transition relation $\delta_M$ by a union of expressions that amount to selecting every pair of tapes that matches one of the transition rules encoded in $\delta^M$:

$$\phi_{succ} := \phi_{witness-succ} \circ (\sigma_{\gamma_1} \cup \cdots \cup \sigma_{\gamma_m}) \circ \text{map}(\pi_s)$$

For instance, if $(q', b, +1) \in \delta(q, a)$, one $\sigma_{\gamma_i}$ is to select the triples $\langle s: \langle S: \langle t: u \triangleright a \triangleleft sv, q: q\rangle, S': \langle t: ub \triangleright s \triangleleft v, q: q'\rangle\rangle, w: \triangleright a \triangleleft s, w': b \triangleright s \triangleleft\rangle \in [\![\phi_{witness-succ}]\!]$. (Details are omitted for lack of space, but it is important to note that the values that we are dealing with are atomic, so we only need equality on atomic values.)

As for $Configs$, $\phi_{succ}$ contains pairs $\langle C, C'\rangle$ of invalid configurations. However, whenever $C$ is a valid configuration, $C'$ is indeed a possible successor configuration on $M$. It follows by induction that, starting from valid configuration $C_{start}$, we will only reach valid configurations via the successor relation $\phi_{succ}$.

**Modeling computations**. Now we are ready to model accepting computations of $M$. Here the problem is the possibly exponential running time. We use a simple recursive divide-and-conquer approach in the spirit of the usual proof

---

[10] Here and later, $\pi_{A_1,\ldots,A_m} := \pi_{A_1} \circ \cdots \circ \pi_{A_m}$.

---

of Savitch's theorem (cf. e.g. [35]). Let $\psi_i$ the pairs of configurations $\langle C, C'\rangle$ such that $C$ is reachable from $C'$ in $2^i$ steps. We define $\psi_i$ as

$$\begin{aligned}
\psi_0 &:= \phi_{succ} \\
\psi_{i+1} &:= \psi_i \circ (\text{id} \times \text{id}) \circ \sigma_{1.C'=2.C} \circ \\
&\quad \text{map}(\langle C: \pi_{1.C}, C': \pi_{2.C'}\rangle)
\end{aligned}$$

Note that the definition of $\psi_{i+1}$ uses $\psi_i$ only once, thus the formula remains computable in LOGSPACE.

There is an accepting computation path of length $2^{K(n)}$ iff there is a pair $\langle C, C'\rangle$ in $\psi_{K(n)}$ such that $C = C_{start}$ and the state of $C'$ is in $F^M$. We can phrase this as

$$\begin{aligned}
\phi_{accept} &:= \Big(\big(\langle 1: C_{start}, 2: \psi_{K(n)}\rangle \circ \text{pairwith}_2 \circ \\
&\quad \sigma_{1=_{mon}2.C} \circ \text{map}(\pi_{2.C'})\big) \times AcceptingConfigs\Big) \circ \\
&\quad \text{map}([1 =_{mon} 2]) \circ \text{flatten}.
\end{aligned}$$

(Again we only employ equality on configurations.)

It is not difficult to see that the $\mathcal{M}_\cup[=_{atomic}]$ query $\phi_{accept}$ constructed is of polynomial size and can be computed in LOGSPACE. The entire problem is formulated as the query (e.g., the input $x$ is constructed from constants and pairs) and $\phi_{accept}$ will not make use of an input value. Thus we have shown that $\mathcal{M}_\cup[=_{atomic}]$ is NEXPTIME-hard with respect to query complexity (i.e., for a fixed database). □

It is not hard to verify that

LEMMA 4.2. *For the construction of the proof of Theorem 4.1, (a)* $|\phi_{accept}| = O(K(n))^2$. *(b) If* $=_{mon}$ *is available as a built-in,* $\phi_{accept}$ *can be defined such that* $|\phi_{accept}| = O(K(n))$.

**Proof Sketch**. We can verify by inspection of the proof of Theorem 4.1 that

$$\begin{aligned}
|Configs| &= O(K(n)), \\
|C_{start}| &= O(K(n)), \\
|AcceptingConfigs| &= |Configs| + O(1) = O(K(n)), \\
|=_{mon}| &= O(K(n)), \\
|\phi_{prepare-succ}| &= O(|Configs|), \\
|\phi_{zoom-in}| &= O(|=_{mon}|), \\
|\phi_{witness-succ}| &= O(|Configs|) \\
&\quad + O(|\phi_{zoom-in}| \cdot K(n)) \\
&= O(K(n)^2), \\
|\phi_{succ}| &= |\phi_{witness-succ}| + O(1) \\
&= O(K(n)^2), \\
|\psi_{K(n)}| &= |\phi_{succ}| + O(K(n) \cdot |=_{mon}|) \\
&= O(K(n)^2), \\
|\phi_{accept}| &= |C_{start}| + |\psi_{K(n)}| \\
&\quad + |AcceptingConfigs| = O(K(n)^2)
\end{aligned}$$

From this it is also clear that (b) if we use built-in $=_{mon}$ operation rather than our defined monotone equality operation, $|\phi_{accept}| = O(K(n))$. □

COROLLARY 4.3. $\mathcal{M}_\cup[=_{mon}]$ *is NETIME-hard under LOGLIN-reductions (query complexity).*

THEOREM 4.4. $\mathcal{M}_\cup[=_{mon}, \text{not}]$ is $TA[2^{O(n)}, O(n)]$-hard under LOGLIN-reductions (query complexity).

**Proof Sketch.** The proof is by a LOGLIN-reduction from $TA[2^{O(n)}, O(n)]$ Turing machine acceptance. Let

$$M = (Q_\exists^M, Q_\forall^M, q_0^M, \delta^M, F^M)$$

be an alternating Turing machine (ATM) that runs in time $2^{O(n)}$ with $O(n)$ alternations on inputs of size $n$.

We simulate the computation of $M$ in $\mathcal{M}_\cup[=_{mon}, \text{not}]$. Each run of $M$ is a tree of configurations of depth $2^{k \cdot n}$, for a suitable constant $k$. We may assume w.l.o.g. that terminating computation paths of $M$ are no longer than $2^{k \cdot n}$, i.e., the depth of the computation tree of $M$ is bounded by $2^{k \cdot n}$.

By Lemma 4.2, if we use a built-in operation $=_{mon}$, the sizes of all formulas of the proof of Theorem 4.1 are linear in the size of $K(n)$. Now, we fix $K(n) = k \cdot n$, for some constant $k$.

We use the formulae $C_{start}$, $Configs$, $AcceptingConfigs$, and $\phi_{succ}$ constructed as described. We define a modified version of $\psi_{k \cdot n}$ which computes the set of computation paths of length *up to* $2^{k \cdot n}$ (this can be realized by adding "stay transitions" $\langle C, C \rangle$, for $C \in Configs$ to $\phi_{succ}$) and where the states of the intermediate configurations are all from $Q_\exists$ if the state of the first configuration is from $Q_\exists$ and are all from $Q_\forall$ otherwise. We can define this as

$$\psi_{i+1} := \psi_i \circ (\text{id} \times \text{id}) \circ \sigma_{1.C'=2.C} \circ$$
$$\sigma_{1.C.q \in Q_\exists^M \Leftrightarrow 2.C.q \in Q_\exists^M} \circ \text{map}(\langle C : \pi_{1.C}, C' : \pi_{2.C'}\rangle).$$

Now that we only need to consider tapes of size $2^{k \cdot n}$, the monad algebra expressions only occupy space $O(n)$, thus so far we have a LOGLIN reduction.

Let the sets of configurations $A_i$ be inductively defined as

$$\begin{aligned}
A_1 &:= \{C \mid \exists C' \ (C, C') \in \psi_{k \cdot n} \wedge \\
&\quad C' \in AcceptingConfigs \wedge C.q \in Q_\exists^M \} \\
A_{i+1} &:= \{C \mid \exists C' \ (C, C') \in \psi_{k \cdot n} \wedge \\
&\quad C' \in (Configs - A_i) \wedge C.q \in Q_\exists^M \Leftrightarrow C'.q \notin Q_\exists^M \}
\end{aligned}$$

Clearly, $C \in A_i$ for odd $i$ means that $C.q \in Q_\exists^M$ and that $C$ is eventually accepting; $C \in A_i$ for even $i$ means that $C.q \in Q_\forall^M$ and that $C$ is not eventually accepting (both via $i$ alternations and in $2^i$ steps).

W.l.o.g., we may assume that $F^M \subseteq Q_\exists^M$. By this assumption $F^M \subseteq A_1$, and thus the final transitions leading to accepting states may be universal, rather than just existential. We will now be somewhat sloppy and assume that the number of alternations $K(n) = O(n)$ we ask for is always odd. This is to keep the argument short, but a slight modification of the construction allows to eliminate the assumption.

Then, $M$ accepts its input precisely if $C_{start}$ is eventually accepting with $K(n)$ alternations, that is, iff $C_{start} \in A_{K(n)}$.

It is not difficult to construct $A_{K(n)}$ in monad algebra. We only remark that difference $A - B$ on sets of nested tuples can be defined using $=_{mon}$ and "not" as

$$\{a \in A \mid \not\exists b \in B \wedge a =_{mon} b\}$$

or, in monad algebra on pair $\langle 1 : A, 2 : B \rangle$,

$$\text{pairwith}_1 \circ \text{flatmap}(\langle a : \pi_1, c : \langle a : \pi_1, B : \pi_2 \rangle \circ \text{pairwith}_B \circ$$
$$\text{flatmap}(a =_{mon} B) \circ \text{not}\rangle \circ \text{pairwith}_c \circ \text{map}(\pi_a))$$

Formula $\phi_{accept}$ is obviously of linear size, and thus the construction in LOGLIN. This concludes our proof. □

Considering again $=_{atomic}$ as a built-in, our LOGSPACE-reduction of the proof of Theorem 4.1 for configurations and $\phi_{succ}$ (with $K(n) = n^k$) in combination with the construction for computations ($A_i$ and $\phi_{accept}$) of the proof of Theorem 4.4 yields

COROLLARY 4.5. $\mathcal{M}_\cup[=_{atomic}, \text{not}]$ is $TA[2^{n^{O(1)}}, n^{O(1)}]$-hard under LOGSPACE-reductions (query complexity).

We can give a more precise lower bound.

THEOREM 4.6. $\mathcal{M}_\cup[=_{atomic}, \text{not}]$ is $TA[2^{O(n)}, O(n)]$-hard under LOGLIN-reductions (query complexity).

**Proof Sketch.** To allow for a query $\phi_{accept}$ of linear size overall, we have to rephrase both $\phi_{witness-succ}$ and $\psi_{K(n)}$ to use our formula defining $=_{mon}$ via $=_{atomic}$ only a constant number of times. We can do this now that we have negation and thus equality of a set of nested tuples available. We only sketch the idea here briefly, but it is the same for the two cases. Rather than testing equality linearly many times, we postpone the testing of equality on pairs of tuples until we have collected all the pairs in a set and we can test equality of them all at once. We demonstrate the idea for $\psi_{K(n)}$. Let

$$\begin{aligned}
\psi_0' &:= \phi_{succ} \circ \text{map}(\langle 1 : \text{id}, 2 : \emptyset \rangle) \\
\psi_{i+1}' &:= (\text{id} \times \text{id}) \circ \sigma_{1.C.q \in Q_\exists^M \Leftrightarrow 2.C.q \in Q_\exists^M} \circ \\
&\quad \text{map}\Big(\langle 1 : \langle C : \pi_{1.1.C}, C' : \pi_{2.1.C'} \rangle, \\
&\quad 2 : \pi_{1.2} \cup \pi_{2.2} \cup \langle 1 : \pi_{1.1.C'}, 2 : \pi_{2.1.C} \rangle \circ \text{sng} \rangle\Big)
\end{aligned}$$

Now we are interested in those pairs of configurations $(c, c')$ s.t. $\langle 1 : \langle C : c, C' : c' \rangle, 2 : S \rangle \in \psi_{K(n)}'$ and for all $\langle 1 : t, 2 : t' \rangle \in S$, $t =_{mon} t'$. We can define this as

$$\psi_{K(n)} := \psi_{K(n)}' \circ \text{map}(\langle 1 : \pi_1, 2 : \pi_2 \circ \text{all-equal}\rangle \circ$$
$$\text{pairwith}_2 \circ \text{map}(\pi_1)) \circ \text{flatten}$$

where all-equal := $\text{map}((1 =_{mon} 2) \circ [\text{not}]) \circ \text{flatten} \circ \text{not}$.

For $\phi_{witness-succ}$, we proceed analogously. We define $\phi_{zoom-in}$ to be a mapping from sets of tuples $\langle s : (C, C'), w : t, w' : t', mbe : S \rangle$ (where $(s : (C, C'), w : t, w' : t')$ is as in the proof of Theorem 4.1 and $S$ is a set of pairs yet to be checked to be equal – $mbe$ is short for "must be equal") to sets of tuples of the same type. We replace e.g. $\sigma_{12 \triangleright 34 \triangleleft} \circ \pi_{12 \triangleright 34 \triangleleft}$ in $\phi_{zoom-in}$ by

$$\text{map}\Big(\langle s : \pi_s, w : \pi_{w.2}, w' : \pi_{w'.2},$$
$$mbe : \langle w : \pi_{w.1}, w' : \pi_{w'.1} \rangle \circ \text{sng} \cup$$
$$\pi_{mbe} \circ \text{map}(\langle w : \pi_{w.1}, w' : \pi_{w'.1} \rangle \circ \text{sng} \cup$$
$$\langle w : \pi_{w.2}, w' : \pi_{w'.2} \circ \text{sng}\rangle) \circ \text{flatten}\rangle\Big)$$

That is, in each such step we add the values that were checked to be equal using $=_{mon}$ in $\phi_{zoom-in}$ – here, for $\sigma_{12 \triangleright 34 \triangleleft} \circ \pi_{12 \triangleright 34 \triangleleft}$, $w.1$ and $w'.1$ – and add them to $mbe$. Before we do that, we split the pairs $(t, t')$ of $mbe$ into their immediate constituents (as shown in the bottom two lines of the monad algebra expression above). This is necessary to assure that all members of $mbe$ are of the same type. However, it has a nice side-effect. By this restructuring, after the last zoom-in step, the members of *mbe* are pairs of *atomic values*, and we actually do not need $=_{mon}$ here at all and can use $=_{atomic}$ instead.

Note that we could not have used this construction in the proof of Theorem 4.1 because now we need negation to check that for each pair $(t, t')$ in *mbe*, $t =_{atomic} t'$. (This can be done using the "all-equal" predicate defined above, with $=_{atomic}$ replacing $=_{mon}$.) □

Since "not" is equivalent to (id = ∅),

COROLLARY 4.7. $\mathcal{M}_\cup[=]$ *is* $TA[2^{O(n)}, n]$-*hard under LOGLIN-reductions (query complexity).*

The queries constructed in our lower bound proofs are from flat relations to flat relations (we may assume this since we actually use no input data value). Since relational algebra is in PSPACE w.r.t. combined complexity (cf. e.g. [3]) and presumably PSPACE ≠ NEXPTIME, it seems unlikely that there is even a PSPACE reduction from $\mathcal{M}_\cup[=]$ on flat relations to relational algebra in the spirit of the Conservativity Theorem of Paredaens and Van Gucht ([36], Theorem 2.5).

## 5. LISTS AND BAGS

In this section, we study the complexity of monad algebra on lists $\mathcal{M}_\cup^{[]}$ and bags $\mathcal{M}_\cup^{\{|\}}$. A formal definition of these languages is beyond the scope of this paper, but see [39, 6, 32] for full formal definitions. We will use the same syntax and operation names as for monad algebra on sets, but now, for instance, ∪ on lists means to append two lists and "flatten" appends the list-typed members of a list in order of appearance. For bags, these operations ignore order but preserve duplicates. Of course, two lists are equal iff they are of the same length and for each $i$, the $i$-th members of the two lists are equal. Two bags are equal iff each member of either bag occurs the same number of times in both bags.

For bags, we will also consider the additional operations "monus" (a powerful version of difference which allows to express arithmetics in monad algebra on bags) and "unique", an operation that eliminates duplicates from bags. In [32] it was shown that adding either of these two operations strictly increases the expressive power of the language (and adding both makes the language yet stronger).

First we again look at data complexity.

PROPOSITION 5.1 (FOLKLORE). $\mathcal{M}_\cup^{\{|\}}[=, monus]$ *and* $\mathcal{M}_\cup^{[]}[=]$ *are in* $TC_0$ *w.r.t. data complexity.*

There is no space to provide a proof for this, but "parsing" and accessing nested data is described in the proof of Proposition 3.2 and implementing the various operations of monad algebra is not difficult. (See also the similar proof that XQuery is in $TC_0$ – Theorem 8.3). It is folklore that the majority gates of $TC_0$ circuits are powerful enough to support the arithmetics required to implement bag operations such as bag difference.

For a result that suggests that this is a good bound,

PROPOSITION 5.2 ([19]). *There are* $\mathcal{M}_\cup^{\{|\}}[=, monus]$ *queries that are not in* $AC_0$.

Regarding query/combined complexity, we can show that

PROPOSITION 5.3. *The languages* $\mathcal{M}_\cup^{\{|\}}[=_{atomic}, not]$ *and* $\mathcal{M}_\cup^{[]}[=_{atomic}, not]$ *are* $TA[2^{O(n)}, O(n)]$-*hard under LOGLIN-reductions (query complexity).*

PROPOSITION 5.4. $\mathcal{M}_\cup^{\{|\}}[=_{atomic}]$ *and* $\mathcal{M}_\cup^{[]}[=_{atomic}]$ *are NEXPTIME-hard w.r.t. query complexity.*

The lower bound proofs for sets work without modifications on lists and bags – we actually do not compare collections except in the definition of $A_i$ of the proof of Theorem 4.4, where we compute differences. But here, we define difference $R - S$ as a *filter* (using "map") that computes those elements of $R$ for which no element of $S$ with the same value exists. For lists, this will preserve order of the elements in $R$ and for bags it will preserve their multiplicities. For the correctness of our reduction, this does not matter, as long as we interpret nonempty collections of type $[\langle\rangle]$ resp. $\{|\langle\rangle|\}$ (possibly with duplicates) as truth and empty collections as falsity.

For the upper bounds,

THEOREM 5.5. $\mathcal{M}_\cup^{\{|\}}[=_{atomic}]$ *and* $\mathcal{M}_\cup^{[]}[=_{atomic}]$ *are in NEXPTIME w.r.t. query complexity.*

THEOREM 5.6. $\mathcal{M}_\cup^{\{|\}}[=_{atomic}, not]$ *and* $\mathcal{M}_\cup^{[]}[=_{atomic}, not]$ *are in* $TA[2^{O(n)}, O(n)]$ *under LOGLIN-reductions (query complexity).*

Here, the proofs of Theorems 3.7 and 3.11 work without modifications for lists and bags. Actually, our encoding using deterministic trees treats collections as lists, and thus preserves both order and multiplicities of members. If only equality on atomic values is available, however, we cannot distinguish between sets, lists, and bags in queries. Finally,

THEOREM 5.7. $\mathcal{M}_\cup^{\{|\}}[=, monus, unique]$ *and* $\mathcal{M}_\cup^{[]}[=]$ *are in EXPSPACE w.r.t. query complexity.*

**Proof Sketch**. The proof is the same as for Theorem 3.13, but now we also have to check deep list and bag equality in EXPSPACE.

Consider the case of $\mathcal{M}_\cup^{\{|\}}[=, monus, unique]$. When we want to check whether two bags identified by the path prefixes $\pi_1$ and $\pi_2$ are equal, we can do this by checking whether for each member of $\pi_1$ or $\pi_2$, its multiplicity in $\pi_1$ is the same as in $\pi_2$. We iteratively compute each root-to-leaf path $\pi_1.i.v$ with prefix $p_1$. For each such path, we write $\pi_1.i$ into an exp-size register. Now, we iterate over $\pi_1$ and count the number of $\pi_1.j$ *equal* to $\pi_1.i$. Then we iterate over $\pi_2$ and count the number of $\pi_2.j$ equal to $\pi_1.i$. If these two counts differ, the values $\pi_1$ and $\pi_2$ are not equal. Now we repeat the same procedure for each root-to-leaf path $\pi_2.i.v$. The values $\pi_1$ and $\pi_2$ are equal if we have not discovered any differing counts. Equality of the values $\pi_1.i$ and $\pi_2.j$ is defined recursively, using the procedure just described, but this is not a problem since the depth of the values is only linear in the size of the query and we thus need only linearly many registers to check deep equality. □

In the remainder of this paper, we will assume that $\mathcal{M}_\cup^{[]}$ has one further operation, "true", which evaluates to $[\langle\rangle]$ (true) on a list if it is nonempty and to $[]$ (false) otherwise. We will use "true" to eliminate duplicate entries from truth values (i.e., from $[\langle\rangle, \ldots, \langle\rangle]$). It is easy to verify that this does not increase the complexity of $\mathcal{M}_\cup^{[]}[=]$, $\mathcal{M}_\cup^{[]}[=_{atomic}]$, or $\mathcal{M}_\cup^{[]}[=_{atomic}, \text{not}]$. For $\mathcal{M}_\cup^{[]}[=]$, in the proof of Theorem 5.7, we only need a small number of exp-size registers to recompute the collection (we can stop early if we find at least one member). In the proofs of the upper bounds of the other two fragments, "true" is a non-operation because for proving a goal, duplicates do not matter.

## 6. CORE XQUERY

We consider the fragment of XQuery with abstract syntax

$$
\begin{aligned}
query &::= \langle a/\rangle \mid \langle a\rangle query \langle/a\rangle \mid query\ query \\
&\mid var \mid var/axis :: \phi \\
&\mid \text{for } var \text{ in } query \text{ return } query \\
&\mid \text{if } cond \text{ then } query \\
&\mid (\text{let } var := query)\ query \\
cond &::= var = var \mid query
\end{aligned}
$$

where $a$ denotes the XML tags, $axis$ the XPath axes[11], $var$ a set of XQuery variables $\$x, \$x_1, \$x_2, \ldots, \$y, \$z, \ldots$, and $\phi$ a *node test* (either a tag name or "*"). We refer to this fragment as *Core XQuery*, or *XQ* for short.

For simplicity, we will work with pure node-labeled unranked ordered trees, and by atomic values, we will refer to leaves (or equivalently, their labels).

XQuery supports several forms of equality. We will not try to use the same syntax (=, eq, or deep_equal) as in the current standards proposal – it is not clear whether the syntax has stabilized. Throughout this paper, equality is by value (that is, by value as a tree rather than by the *yield* of strings at leaf nodes of the tree). We will write $=_{deep}$ and $=_{atomic}$ for deep and atomic equality, respectively. We will use = for statements that apply to both forms of equality.

The semantics of *XQ* is given in Figure 3. As for $\mathcal{M}_\cup^{[]}$, $\cup$ here denotes list concatenation; $<_{doc}$ is the depth-first left-to-right traversal order through the tree, $\chi^t$ is the axis relation $\chi$ on tree $t$, $\text{lab}_*^t$ is true on all nodes of $t$, and $\text{lab}_a^t$, for $a$ a tag name, is true on those nodes of $t$ labeled $a$. All *XQ* queries evaluate to lists of nodes; however, we assume that *XQ* variables always bind to single nodes rather than lists; to assure this, we require that for expressions "(let $\$x_{k+1} := \alpha$) $\beta$", $\alpha$ is either of the form $\langle a/\rangle$ or $\langle a\rangle\ \alpha_0\ \langle/a\rangle$" (i.e., this gives a easy syntactic condition that $\alpha$ always evaluates to a singleton list). This semantics is (observationally) consistent with [44] restricted to Core XQuery.

In our definition of the syntax of Core XQuery, we have been economical with operators introduced. Since conditions are true iff they evaluate to a nonempty collection,

$$
\begin{aligned}
\phi \text{ or } \psi &:= \phi\ \psi \\
\phi \text{ and } \psi &:= \text{if } \phi \text{ then } \psi \\
\text{some }\$x \text{ in } \alpha \text{ satisfies } \phi &:= \text{for }\$x \text{ in } \alpha \text{ return } \phi
\end{aligned}
$$

Using deep equality, we can define negation,

$$\text{not } \phi := \big((\langle a\rangle\{\text{if } \phi \text{ then } \langle b/\rangle\}\langle/a\rangle) =_{deep} \langle a/\rangle\big).$$

[11]For simplicity, particularly of the following semantics, we will only consider the child and the descendant axis; but all complexity upper bounds will hold for all XPath axes. Our theorems will refer to "all axes", but proofs will assume that this notion refers just to the two axes child and descendant.

$$
\begin{aligned}
[\![\langle a/\rangle]\!]_k(\vec{e}) &:= [\langle a/\rangle] \\
[\![\langle a\rangle\alpha\langle/a\rangle]\!]_k(\vec{e}) &:= [\langle a\rangle[\![\alpha]\!](\vec{e})\langle/a\rangle] \\
[\![\alpha\ \beta]\!]_k(\vec{e}) &:= [\![\alpha]\!](\vec{e}) \cup [\![\beta]\!](\vec{e}) \\
[\![\text{for }\$x_{k+1} \text{ in } \alpha & \\
\quad\text{return } \beta]\!]_k(\vec{e}) &:= \text{let } l = [\![\alpha]\!]_k(\vec{e}), n = |l|; \\
& \quad\text{return } \bigcup_{1 \leq i \leq n} [\![\beta]\!]_{k+1}(\vec{e}, l_i) \\
[\![(\text{let }\$x_{k+1} := \alpha)\ \beta]\!]_k(\vec{e}) &:= [\![\beta]\!]_{k+1}(\vec{e}, [\![\alpha]\!]_k(\vec{e})) \\
[\![\$x_i]\!]_k(t_1, \ldots, t_n) &:= t_i \\
[\![\$x_i/\chi :: \phi]\!]_k(\vec{e}) &:= \text{list of nodes } v \text{ of tree } t \text{ s.t.} \\
& \quad [\![\$x_i]\!]_k(\vec{e}) = [t] \land \\
& \quad \chi^t(\text{root}^t, v) \land \text{lab}_\phi^t(v) \\
& \quad \text{in order } <_{doc} \\
[\![\text{if } \gamma \text{ then } \alpha]\!]_k(\vec{e}) &:= \text{if } [\![\gamma]\!]_k(\vec{e}) \text{ then } [\![\alpha]\!]_k(\vec{e}) \text{ else } [] \\
[\![\$x_i = \$x_j]\!]_k(e_1, \ldots, e_k) &:= \text{if } e_i = e_j \text{ then } [\langle yes/\rangle] \text{ else } []
\end{aligned}
$$

**Figure 3: Semantics of Core XQuery.**

Conditions "every $\$x$ in $\alpha$ satisfies $\phi$" can be defined using "not" and "some". We can even assume that expressions of the form $var/path$ are supported, where $path$ is any expression in navigational XPath (aka. Core XPath [16, 17]).

It is clear that

PROPOSITION 6.1. *Let $\mathbf{X}$ be a set of operations and axes. Then, each $XQ[=_{deep}, \text{not}, \text{every}, \mathbf{X}]$ query can be translated in LOGLIN into an equivalent $XQ[=_{deep}, \mathbf{X}]$ query and each $XQ[\text{and}, \text{or}, \text{some}, \mathbf{X}]$ query can be translated in LOGLIN into an equivalent $XQ[\mathbf{X}]$ query.*

Next, we provide mappings between Core XQuery (using only the child axis) and monad algebra on lists. These show the equivalence of these languages up to representation issues, but our main aim is to provide reductions for the study of the complexity of XQuery.

### Translation from Core XQuery to $\mathcal{M}_\cup^{[]}$

We recursively map the data tree $T$ to a complex value $C(T)$ as follows: Each tree node with label $t$ and children $v_1, \ldots, v_n$ is mapped to a tuple

$$\langle \text{label} : t, \text{children} : \{C(v_1), \ldots, C(v_n)\}\rangle.$$

Modulo representation issues captured in the tree translation function $C$, there is an equivalent monad algebra query for each $XQ[=, \text{child}]$ query, for = either $=_{deep}$ or $=_{atomic}$:

THEOREM 6.2. *There is a mapping*

$$MA : XQ[=, child, \text{not}] \to \mathcal{M}_\cup^{[]}[=, \text{not}]$$

*such that for each $XQ[=, child, \text{not}]$ query $Q$,*

1. *for any XML document tree $T$,*

   $$[C(Q(T))] = MA(Q)(\{\langle N : \$ROOT, V : [C(T)]\rangle\}),$$

2. *$MA(Q)$ can be computed in space $O(\log |Q|)$, and*

3. *$|MA(Q)| = O(|Q|)$.*

$$MA \quad : \quad XQ[=, \text{child}, \text{not}] \to \{\langle N : \text{varname}, V : \tau\rangle\} \to \tau'$$

$$
\begin{aligned}
MA(\alpha\ \beta) &:= MA(\alpha) \cup MA(\beta) \\
MA(\langle a/\rangle) &:= \langle \text{label} : a, \text{children} : []\rangle \circ \text{sng} \\
MA(\langle a\rangle\alpha\langle /a\rangle) &:= \langle \text{label} : a, \text{children} : MA(\alpha)\rangle \circ \text{sng} \\
MA(\$x_i) &:= \pi_i \\
MA(\$x_i/t) &:= \pi_i \circ \text{flatmap}(\pi_{\text{children}} \circ \sigma_{\text{label}=t}) \\
MA(\text{for } \$x \text{ in } \alpha \text{ return } \beta) &:= \langle 1 : \text{id}, 2 : MA(\alpha)\rangle \circ \text{pairwith}_2 \circ \text{flatmap}((\pi_1 \cup (\langle N : \$x, V : \pi_2\rangle \circ \text{sng})) \circ MA(\beta)) \\
MA((\text{let } \$x := \alpha)\ \beta) &:= \langle 1 : \text{id}, 2 : MA(\alpha)\rangle \circ \text{pairwith}_2 \circ \text{flatmap}((\pi_1 \cup (\langle N : \$x, V : \pi_2\rangle \circ \text{sng})) \circ MA(\beta)) \\
MA(\text{if } \alpha \text{ then } \beta) &:= \langle 1 : \text{id}, 2 : MA(\alpha) \circ \text{true}\rangle \circ \text{pairwith}_2 \circ \text{flatmap}(\pi_1 \circ MA(\beta)) \\
\\
MA(\text{not } \alpha) &:= MA(\alpha) \circ \text{map}(\langle\rangle) \circ \text{not} \\
MA(\$x = \$y) &:= \langle 1 : \sigma_{N=\$x}, 2 : \sigma_{N=\$y}\rangle \circ \text{pairwith}_1 \circ \text{flatmap}(\text{pairwith}_2) \circ \sigma_{1.V=2.V}
\end{aligned}
$$

**Figure 4: Mapping from** $XQ[=, \text{child}, \text{not}]$ **to** $\mathcal{M}_\cup^{[]}[=, \text{not}]$.

**Proof Sketch.** It is easy to verify that the function $MA$ of Figure 4 satisfies conditions (1) to (3) of our theorem. □

For atomic equality, $\sigma_{1.V=2.V}$ in the definition of $MA$ is to be implemented as $\sigma_{1.V.\text{label}=_{atomic}2.V.\text{label}}$. Note that on a $XQ[=, \text{child}]$ query $Q$, $MA(Q)$ is a $\mathcal{M}_\cup^{[]}[=]$ query.

*Translation from $\mathcal{M}_\cup^{[]}$ to Core XQuery*

Let $T$ be the following canonical translation from complex values to trees:

$$T(\langle A_1 : v_1, A_2 : v_2\rangle) = \\ \langle tup\rangle\langle a_1\rangle T(v_1)\langle /a_1\rangle\langle a_2\rangle T(v_2)\langle /a_2\rangle\langle /tup\rangle$$

$$T(\{v_1, \ldots, v_n\}) = \langle list\rangle T(v_1) \ldots T(v_n)\langle /list\rangle$$

Note that $T$ is not the inverse of the mapping $C$ that we introduced above to map from XML trees to complex values constructed from tuples and lists.

Let $\mathcal{M}_\cup^{[],(\cdot,\cdot)}$ denote the monad algebra queries on lists and pairs (rather than on tuples of arbitrary arity). For both $=_{deep}$ and $=_{atomic}$, we have

THEOREM 6.3. *There is a mapping*

$$XQ : \mathcal{M}_\cup^{[]}[=] \to XQ[=, \text{child}]$$

*such that for each $\mathcal{M}_\cup^{[]}[=]$ query $Q$,*

1. *for any complex value $v$,*
   $$T(Q(v)) = XQ(Q)(\$ROOT)\big(T(v)\big),$$

2. *$XQ(Q)$ can be computed in space $O(\log|Q|)$, and*

3. *If $Q$ is a $\mathcal{M}_\cup^{[],(\cdot,\cdot)}$ query, $|XQ(Q)| = O(|Q|)$.*

**Proof Sketch.** It is easy to verify that the mapping of Figure 5 satisfies (1) to (3). □

## 7. COMBINED COMPLEXITY OF XQ

Theorems 6.2 and 6.3 provide LOGLIN-reductions back and forth between queries in monad algebra on lists and $XQ$. Note that each XML document is an XQuery. We can compose query and data into a single query (certainly in LOGLIN). All the complexity results of this section involve complexity classes that are closed under LOGLIN-reductions. Thus combined complexity for Core XQuery with composition will never be harder than query complexity, and we only need to study the latter.

COROLLARY 7.1. *With respect to query complexity,*

- *$XQ[=_{deep}, \text{child}]$ is $TA[2^{O(n)}, O(n)]$-hard under LOGLIN-reductions and in EXPSPACE;*

- *$XQ[=_{atomic}, \text{child}, \text{not}]$ is $TA[2^{O(n)}, O(n)]$-complete under LOGLIN-reductions;*

- *$XQ[=_{atomic}, \text{child}]$ is NEXPTIME-complete.*

**Proof.** This follows immediately from the results of Section 5 on $\mathcal{M}_\cup^{[]}$ and, for the upper bounds, the LOGSPACE-reduction from $XQ[=, \text{child}]$ to $\mathcal{M}_\cup^{[]}[=]$ of Theorem 6.2 resp., for the lower bounds, the LOGLIN-reduction from $\mathcal{M}_\cup^{[]}[=]$ to $XQ[=, \text{child}]$ of Theorem 6.3. □

The complexity classes in which our XQuery evaluation problems reside are large enough that minor extensions, such as supporting all XPath axes, do not matter.

THEOREM 7.2. *W.r.t. combined complexity,*

- *$XQ[=_{deep}, \text{all axes}]$ is in EXPSPACE,*

- *$XQ[=_{atomic}, \text{all axes}, \text{not}]$ is in $TA[2^{O(n)}, O(n)]$, and*

- *$XQ[=_{atomic}, \text{all axes}]$ is in NEXPTIME.*

A proof of this is not difficult but is beyond the scope of this paper and will be given in its long version. For the EXPSPACE bound, for instance, the proof idea is that no value computable by an $XQ$ query is greater than of doubly exponential size in the size of the input (see also Proposition 3.4 for the analogous fact for monad algebra on lists). Thus we can use "addresses" of singly exponential size to refer to subtrees of an intermediate result. While we cannot store the subtree as a whole, we can always recompute it from these addresses and the query. There are only linearly many variables in an XQuery, so we need only linearly

$$\begin{aligned}
XQ(\langle A_1 : f_1, \ldots A_k : f_k\rangle)(\$x) &:= T(\langle A_1 : XQ(f_1)(\$x), \ldots, A_k : XQ(f_k)(\$x)\rangle) \\
XQ(\pi_i)(\$x) &:= \{\$x/a_i/*\} \\
XQ(\text{sng})(\$x) &:= \langle list\rangle\{\$x\}\langle /list\rangle \\
XQ(f \circ g)(\$x) &:= (\text{let } \$y := XQ(f)(\$x)) \; XQ(g)(\$y) \\
XQ(\text{map}(f))(\$x) &:= \langle list\rangle\{\text{for } \$y \text{ in } \$x/* \text{ return } XQ(f)(\$y)\}\langle /list\rangle \\
XQ(\text{id})(\$x) &:= \$x \\
XQ(\text{flatten})(\$x) &:= \langle list\rangle\{\$x/list/*\}\langle /list\rangle \\
XQ(\text{pairwith}_i)(\$x) &:= \langle list\rangle\{\text{for } \$y \text{ in } \$x/a_i/\text{list}/* \text{ return } \langle tup\rangle \\
&\qquad \langle a_1\rangle\{\$x/a_1/*\}\langle /a_1\rangle \ldots \langle a_{i-1}\rangle\{\$x/a_{i-1}/*\}\langle /a_{i-1}\rangle \\
&\qquad \langle a_i\rangle\{\$y\}\langle /a_i\rangle \\
&\qquad \langle a_{i+1}\rangle\{\$x/a_{i+1}/*\}\langle /a_{i+1}\rangle \ldots \langle a_k\rangle\{\$x/a_k/*\}\langle /a_k\rangle \\
&\qquad \langle /tup\rangle\}\langle /list\rangle \\
XQ(f \cup g)(\$x) &:= \langle list\rangle\{(XQ(f)(\$x))/*\}\{(XQ(g)(\$x))/*\}\langle /list\rangle \\
XQ(\sigma_{A_i = A_j})(\$x) &:= \{\text{ for } \$y \text{ in } \$x/* \text{ return } \{\text{ if } (\$x/a_i/* = \$x/a_j/*) \text{ then } \$x \} \} \\
XQ(c)(\$x) &:= \langle c/\rangle
\end{aligned}$$

**Figure 5:** Mapping from $\mathcal{M}_\cup^{[]}$ to $XQ[=_{deep}, \text{child}]$.

many exp-size registers to evaluate the query. This yields an EXPSPACE algorithm.

Note that this EXPSPACE algorithm is quite robust and allows to add a number of XQuery features that we excluded from $XQ$, such as counting, (document position) arithmetics, and duplicate elimination.

## 8. DATA COMPLEXITY OF XQ

By Proposition 5.1, monad algebra is in LOGSPACE with respect to data complexity. Since LOGSPACE is closed under composition (cf. [35]) and the mapping $C$ is clearly in LOGSPACE,

COROLLARY 8.1. $XQ[=_{deep}, \text{child}]$ *is in LOGSPACE w.r.t. data complexity.*

We can improve on this result.

The data complexity of XQuery is so low that we need to be careful about how the XML data is represented. We distinguish the cases of representation by a DOM tree (i.e., a pointer structure) and representation by a string (an XML document). As we show, the complexity is (presumably) slightly lower on strings than on trees, even though the former require parsing the input. It turns out that the complexity bounds are precisely the same as for XPath [17, 37].

THEOREM 8.2. $XQ[=_{deep}, \text{all axes}]$ *is LOGSPACE-complete under $NC_1$-reductions (data complexity) if the input is given as a DOM tree.*

**Proof Sketch.** Let us assume that for a given XQuery, already the result of all its subqueries are given as input. Then we can evaluate the query in LOGSPACE because all the space we need is a fixed number of log-sized registers for the variables of the query. (We only need a logarithmic number of bits to store a node id of the input).

A fixed query consist of a fixed number of compositions. Since LOGSPACE is closed under compositions (cf. [35]), we can compose the algorithm just discussed for the individual subqueries into a single LOGSPACE algorithm that intuitively computes the query by precomputing its subqueries bottom-up (w.r.t. the syntax tree of the query).

LOGSPACE-hardness follows from the fact that directed tree reachability is LOGSPACE-complete under $NC_1$-reductions [8] and directed tree reachability (checking whether node $w$ is reachable from node $v$ in tree $t$) can be easily encoded by mapping $t$ to a XML tree in which only $v$ has label "v" and only $w$ has label "w". Then the query /descendant::v/descendant::w tests reachability of $w$ from $v$. □

THEOREM 8.3. $XQ[=_{deep}, \text{all axes}]$ *is in $TC_0$ w.r.t. data complexity if the XML input is given as a character string.*

**Proof Sketch.** We show a stronger result, that every Core XQuery expression can be encoded as a TC0 reduction that transforms the input data into the query result.

By $FOM$, we denote first-order logic extended with majority quantifiers $M$ [4]. A formula $My\,\phi(\vec{x}, y)$ is true if $\phi(\vec{x}, y)$ is true for more than half of the positions $y$ of the input. It is known that $TC_0$ is equivalent to the class of languages recognizable using FOM sentences [4].

The reduction is encoded in FOM. A FOM reduction [4] is a set of FOM formulae, consisting of a formula "size" s.t. size($s$) iff the size of the string is $s$ and a formula $\text{pos}_a$ for each $a \in \Sigma$ s.t. $\text{pos}_a(i)$ iff the $i$-th symbol of the string is "a". It is known [4] that FOM can express predicates $x = y + z$ and $x = \#y\,\phi(y)$, such that $x$ is the number of positions $y$ for which $\phi(y)$ holds. We use $\Sigma\{y \mid \phi(\vec{x}, y)\}$ as a shortcut for $\#u\,\exists y : \phi(\vec{x}, y) \wedge 1 \leq u \leq y$.

We will assume the document to be encoded using an alphabet of opening and matching closing tags. For the sake of simplicity of this proof, but without loss of generality, we will assume that base values (e.g. strings) are encoded as trees. The input will be a well-formed sequence of opening and closing tags. We will identify nodes by the position of their opening tag in the (input) string. The input is represented using a predicate size$[\![\$ROOT]\!]$ s.t. size$[\![\$ROOT]\!](n)$ iff $n$ is the size of the input and predicates $\text{pos}_a[\![\$ROOT]\!]$ s.t. $\text{pos}_a[\![\$ROOT]\!](i)$ iff the $i$-th symbol of the input is "a".

$$\begin{aligned}
\text{size}[\![\langle a/\rangle]\!]_k(\vec{e}, s) &:\Leftrightarrow 2\\
\text{pos}_l[\![\langle a/\rangle]\!]_k(\vec{e}, i) &:\Leftrightarrow (i = 1 \Rightarrow l = \langle a \rangle) \wedge (i = 2 \Rightarrow l = \langle /a \rangle)\\[4pt]
\text{size}[\![\langle a \rangle \alpha \langle /a \rangle]\!]_k(\vec{e}, s) &:\Leftrightarrow \exists s' : \text{size}[\![\alpha]\!]_k(\vec{e}, s') \wedge s = 2 + s'\\
\text{pos}_l[\![\langle a \rangle \alpha \langle /a \rangle]\!]_k(\vec{e}, i) &:\Leftrightarrow \exists s : \text{size}[\![\langle a \rangle \alpha \langle /a \rangle]\!]_k(\vec{e}, s) \wedge\\
&\quad (i = 1 \Rightarrow l = \langle a \rangle) \wedge\\
&\quad (1 < i < s \Rightarrow \text{pos}_l[\![\alpha]\!]_k(\vec{e}, i-1)) \wedge\\
&\quad (i = s \Rightarrow l = \langle /a \rangle)\\[4pt]
\text{size}[\![\alpha\,\beta]\!]_k(\vec{e}, s) &:\Leftrightarrow \exists s_1 \exists s_2 : s = s_1 + s_2 \wedge \text{size}[\![\alpha]\!]_k(\vec{e}, s_1) \wedge \text{size}[\![\beta]\!]_k(\vec{e}, s_2)\\
\text{pos}_l[\![\alpha\,\beta]\!]_k(\vec{e}, i) &:\Leftrightarrow \exists s : \text{size}[\![\alpha]\!]_k(\vec{e}, s) \wedge\\
&\quad (i \leq s \Rightarrow \text{pos}_l[\![\alpha]\!]_k(\vec{e}, i)) \wedge\\
&\quad (i > s \Rightarrow \exists i' : i' = i - s \wedge \text{pos}_l[\![\beta]\!]_k(\vec{e}, i'))\\[4pt]
\text{size}[\![(\text{let } \$x_{k+1} := \alpha)\ \beta]\!]_k(\vec{e}, s) &:\Leftrightarrow \text{size}[\![\beta]\!]_{k+1}(\vec{e}, 1, s)\\
\text{pos}_l[\![(\text{let } \$x_{k+1} := \alpha)\ \beta]\!]_k(\vec{e}, i) &:\Leftrightarrow \text{pos}_l[\![\beta]\!]_{k+1}(\vec{e}, 1, i)\\[4pt]
\text{size}[\![\text{for } \$x_{k+1} \text{ in } \alpha \text{ return } \beta]\!]_k(\vec{e}, s) &:\Leftrightarrow s = \Sigma\{s' \mid \exists j : \text{item}[\![\alpha]\!]_k(\vec{e}, j) \wedge \text{size}[\![\beta]\!]_{k+1}(\vec{e}, j, s')\}\\
\text{pos}_l[\![\text{for } \$x_{k+1} \text{ in } \alpha \text{ return } \beta]\!]_k(\vec{e}, i) &:\Leftrightarrow \exists s \exists j_0 : s = \Sigma\{s' \mid \exists j : j < j_0 \wedge \text{item}[\![\alpha]\!]_k(\vec{e}, j) \wedge \text{size}[\![\beta]\!]_{k+1}(\vec{e}, j, s')\} \wedge\\
&\quad \exists s' : \text{item}[\![\alpha]\!]_k(\vec{e}, s+1) \wedge \text{size}[\![\beta]\!]_{k+1}(\vec{e}, s+1, s')\} \wedge\\
&\quad s < i \leq s + s' \wedge \text{pos}_l[\![\beta]\!]_{k+1}(\vec{e}, s+1, i-s)\\[4pt]
\text{size}[\![\$x_i / \chi :: a]\!]_k(\vec{e}, s) &:\Leftrightarrow s = \Sigma\{j' - j + 1 \mid \text{axis}_\chi[\![\$x_i]\!]_k(\vec{e}, 1, j) \wedge \text{node}[\![\$x_i]\!]_k(\vec{e}, j, j')\}\\
\text{pos}_a[\![\$x_i / \chi :: a]\!]_k(\vec{e}, i) &:\Leftrightarrow \exists s \exists j_0 : s = \Sigma\{j' - j + 1 \mid \exists j : j < j_0 \wedge\\
&\quad \text{axis}_\chi[\![\$x_i]\!]_k(\vec{e}, 1, j) \wedge \text{node}[\![\$x_i]\!]_k(\vec{e}, j, j')\} \wedge\\
&\quad \exists s' \exists j'_0 : \text{axis}_\chi[\![\$x_i]\!]_k(\vec{e}, 1, j_0) \wedge \text{node}[\![\$x_i]\!]_k(\vec{e}, j_0, j'_0)\} \wedge s' = j'_0 - j_0 + 1 \wedge\\
&\quad s < i \leq s + s' \wedge \text{pos}_l[\![\$x_i]\!]_k(\vec{e}, i - s + j_0 - 1)\\[4pt]
\text{size}[\![\$x_i]\!]_k(x_1, \ldots, x_k, s) &:\Leftrightarrow \exists j : \text{node}[\![\text{expr}(\$x_i)]\!]_{i-1}(x_1, \ldots, x_{i-1}, x_i, j) \wedge\\
&\quad s = j - x_i + 1\\
\text{pos}_l[\![\$x_i]\!]_k(x_1, \ldots, x_k, i) &:\Leftrightarrow \text{pos}_l[\![\text{expr}(\$x_i)]\!]_{i-1}(x_1, \ldots, x_{i-1}, x_i + i - 1)\\[4pt]
\text{size}[\![\$root]\!]_k(\vec{e}, s) &:\Leftrightarrow \text{size}(s)\\
\text{pos}_l[\![\$root]\!]_k(\vec{e}, i) &:\Leftrightarrow \text{pos}_l(i)\\[4pt]
\text{size}[\![\text{if } \Phi \text{ then } \alpha]\!]_k(\vec{e}, s) &:\Leftrightarrow (\text{cond}[\![\Phi]\!]_k(\vec{e}) \Rightarrow \text{size}[\![\alpha]\!]_k(\vec{e}, s)) \wedge ((\neg\text{cond}[\![\Phi]\!]_k(\vec{e})) \Rightarrow s = 0)\\
\text{pos}_l[\![\text{if } \Phi \text{ then } \alpha]\!]_k(\vec{e}, i) &:\Leftrightarrow \text{pos}_l[\![\alpha]\!]_k(\vec{e}, i)\\[4pt]
\text{cond}[\![\$x_i =_{deep} \$x_j]\!]_k(\vec{e}) &:\Leftrightarrow \exists s : \text{size}[\![\$x_i]\!]_k(\vec{e}, s) \wedge \text{size}[\![\$x_j]\!]_k(\vec{e}, s) \wedge\\
&\quad \forall p : 1 \leq p \leq s \Rightarrow \bigwedge_{l \in \Sigma} \text{pos}_l[\![\$x_i]\!](\vec{e}, p) \Leftrightarrow \text{pos}_l[\![\$x_j]\!](\vec{e}, p)\\[4pt]
\text{axis}_{\text{descendant}}[\![\alpha]\!]_k(\vec{e}, i, j) &:\Leftrightarrow \exists i', j'\ \text{node}[\![\alpha]\!]_k(\vec{e}, i, i') \wedge \text{node}[\![\alpha]\!]_k(\vec{e}, j, j') \wedge i < j \wedge j' < i'\\
\text{axis}_{\text{child}}[\![\alpha]\!]_k(\vec{e}, i, j) &:\Leftrightarrow \exists i', j'\ \text{node}[\![\alpha]\!]_k(\vec{e}, i, i') \wedge \text{node}[\![\alpha]\!]_k(\vec{e}, j, j') \wedge\\
&\quad i < j \wedge j' < i' \wedge \nexists l, l' : \text{node}[\![\alpha]\!]_k(\vec{e}, l, l') \wedge i < l < j \wedge j' < l' < i'.\\[4pt]
\text{item}[\![\alpha]\!]_k(\vec{e}, i) &:\Leftrightarrow \exists i' : \text{node}[\![\alpha]\!]_k(\vec{e}, i, i') \wedge \nexists j, j' : \text{node}[\![\alpha]\!]_k(\vec{e}, j, j') \wedge j < i \wedge i' < j'
\end{aligned}$$

**Figure 6: FOM encoding of the Core XQuery evaluation problem.**

Let

$$\mathrm{node}[\![\alpha]\!]_k(\vec{e}, i, i') :\Leftrightarrow$$
$$\bigvee_{\langle a \rangle \in \Sigma} \mathrm{pos}_{\langle a \rangle}[\![\alpha]\!]_k(\vec{e}, i) \wedge \mathrm{pos}_{\langle /a \rangle}[\![\alpha]\!]_k(\vec{e}, i') \wedge$$
$$\#u(i < u < i' \wedge \mathrm{pos}_{\langle a \rangle}[\![\alpha]\!]_k(\vec{e}, u)) =$$
$$\#u(i < u < i' \wedge \mathrm{pos}_{\langle /a \rangle}[\![\alpha]\!]_k(\vec{e}, u))$$

That is, $\mathrm{node}[\![\alpha]\!]_k(\vec{e}, i, i')$ is true iff $i$ and $i'$ are the positions of a opening tag and a matching closing tag. Since we may assume that the document is well-formed, a sufficient condition for $i'$ being the closing tag matching the opening tag $i$ is that the number of opening tags $\langle a \rangle$ between $i$ and $i'$ is the same as the number of closing tags $\langle /a \rangle$ (i.e., other tags do not have to be considered).

Now, our $XQ[=_{deep}]$ query $Q$ can be encoded by FOM formulas $\mathrm{pos}_l[\![Q]\!]_k$ and $\mathrm{size}[\![Q]\!]_k$ as shown in Figure 6 (for most $XQ$ constructs). (Because of space limitations, we will only be able to provide the full construction in the long version of this paper, but it is straightforward to supplement the construction for the remaining operations.) Note that in $(\mathrm{pos}_l[\![\alpha]\!]_k(\vec{e}, i)$, $(\mathrm{size}[\![\alpha]\!]_k(\vec{e}, s)$, $\vec{e}$ denotes the environment for $k$ variables, indicating positions/nodes assigned to known variables.

Let the *defining expression* for a variable \$x, $expr(\$x)$, be $\alpha$ if \$x is introduced in an $XQ$ expression "for \$x in $\alpha$ return $\beta$" or "let \$x := $\alpha$".

To get an intuition for the reduction of Figure 6, consider again our $XQ$ semantics definition of Figure 3. There, the environments $\vec{e}$ are tuples of valuations of XQuery variables (i.e., trees). Consider the minor reformulation of the semantics that we get if we assume that the value of each variable $\$x_i$ in an environment is an integer that indicates the position of the starting tag of the node it binds to in the value of $defexpr(\$x)$. To get a correct semantics definition along these lines, we just have to set

$$[\![\$x_i]\!]_k(t_1, \ldots, t_n) := [\![expr(\$x_i)]\!]_{i-1}(t_1, \ldots, t_{i-1})$$
$$[\![(\text{let } \$x_{k+1} := \alpha) \beta]\!]_k(\vec{e}) := [\![\beta]\!]_{k+1}(\vec{e}, 1)$$

and an analogous definition for "for"-expressions (which sets the value in the environment to the start index of node $v$ in $[\![expr(\$x_i)]\!]_{i-1}(t_1, \ldots, t_{i-1})$ rather than the node itself). Figure 3 now shows a direct encoding of this altered semantics in FOM.

Considering the problem of deciding whether the root node of the query result has a child as the decision problem for query evaluation, we encode it in FOM (for query $Q$) as $\exists s : size[\![Q]\!]_1(1, s) \wedge s > 2$. □

## Acknowledgments

I thank Michael Benedikt, Peter Buneman, Leonid Libkin, Nicole Schweikardt, Val Tannen, and Andrei Voronkov for their remarks that helped to improve this paper.

## 9. REFERENCES

[1] S. Abiteboul and C. Beeri. "The Power of Languages for the Manipulation of Complex Values". *VLDB J.*, **4**(4):727–794, 1995.

[2] S. Abiteboul and G. G. Hillebrand. "Space Usage in Functional Query Languages". In *Proc. of the 5th International Conference on Database Theory (ICDT)*, pages 439–454, 1995.

[3] S. Abiteboul, R. Hull, and V. Vianu. *Foundations of Databases*. Addison-Wesley, 1995.

[4] D. A. M. Barrington, N. Immerman, and H. Straubing. "On Uniformity within NC1". *Journal of Computer and System Sciences*, **41**(3):274–306, 1990.

[5] L. Berman. "The Complexity of Logical Theories". *Theoretical Comput. Sci.*, **11**:216–224, 1980.

[6] P. Buneman, S. A. Naqvi, V. Tannen, and L. Wong. "Principles of Programming with Complex Objects and Collection Types". *Theor. Comput. Sci.*, **149**(1):3–48, 1995.

[7] A. K. Chandra, D. C. Kozen, and L. J. Stockmeyer. "Alternation". *Journal of the ACM*, 28:114–133, 1981.

[8] S. A. Cook and P. McKenzie. "Problems Complete for Deterministic Logarithmic Space". *J. Algorithms*, **8**:385–394, 1987.

[9] E. Dantsin, T. Eiter, G. Gottlob, and A. Voronkov. "Complexity and Expressive Power of Logic Programming". *ACM Computing Surveys*, **33**(3):374–425, Sept. 2001.

[10] E. Dantsin and A. Voronkov. "Complexity of Query Answering in Logic Databases with Complex Values". In *Proc. LFCS'97, LNCS 1234*, pages 56–66, 1997.

[11] E. Dantsin and A. Voronkov. Expressive Power and Data Complexity of Query Languages for Trees and Lists. In *Proceedings of the 19th ACM SIGACT-SIGMOD-SIGART Symposium on Principles of Database Systems (PODS'00)*, pages 157–165, Dallas, Texas, USA, 2000. ACM Press.

[12] M. Fernandez, J. Siméon, and P. Wadler. An Algebra for XML Query. In *Proc. FSTTCS 2000*, LNCS 1974, pages 11–45. Springer-Verlag, 2000.

[13] M. F. Fernandez and J. Siméon. "Building an Extensible XQuery Engine: Experiences with Galax (Extended Abstract)". In *Proc. XSYM*, pages 1–4, 2004.

[14] J. Ferrante and C. Rackoff. "A Decision Procedure for the First Order Theory of Real Addition with Order". *SIAM J. Comput.*, **4**(1):69–76, 1975.

[15] D. Florescu, C. Hillery, D. Kossmann, P. Lucas, F. Riccardi, T. Westmann, M. J. Carey, A. Sundararajan, and G. Agrawal. "The BEA/XQRL Streaming XQuery Processor". In *Proceedings of the 29th International Conference on Very Large Data Bases (VLDB)*, pages 997–1008, 2003.

[16] G. Gottlob, C. Koch, and R. Pichler. "Efficient Algorithms for Processing XPath Queries". In *Proceedings of the 28th International Conference on Very Large Data Bases (VLDB)*, pages 95–106, Hong Kong, China, 2002.

[17] G. Gottlob, C. Koch, and R. Pichler. "The Complexity of XPath Query Evaluation". In *Proceedings of the 22nd ACM SIGACT-SIGMOD-SIGART Symposium on Principles of Database Systems (PODS)*, pages 179–190, San Diego, California, 2003.

[18] R. Greenlaw, H. J. Hoover, and W. L. Ruzzo. *Limits to Parallel Computation: P-Completeness Theory*. Oxford University Press, 1995.

[19] S. Grumbach, L. Libkin, T. Milo, and L. Wong. "Query Languages for Bags – Expressive Power and Complexity". *SIGACT News*, **27**(2):30–44, 1996.

[20] S. Grumbach and T. Milo. "Towards Tractable Algebras for Bags". In *Proceedings of the 12th ACM SIGACT-SIGMOD-SIGART Symposium on Principles of Database Systems (PODS'93)*, pages 49–58, 1993.

[21] S. Grumbach and V. Vianu. "Tractable Query Languages for Complex Object Databases". In *Proceedings of the 10th ACM SIGACT-SIGMOD-SIGART Symposium on Principles of Database Systems (PODS'91)*, pages 315–327, 1991.

[22] J. Hartmanis, P. M. Lewis II, and R. E. Stearns. "Hierarchies of Memory Limited Computations". In *Proc. Sixth Annual IEEE Symposium on Switching Circuit Theory and Logical Design*, pages 179–190, 1965.

[23] J. Hidders, J. Paredaens, R. Verkammen, and S. Demeyer. "A Light but Formal Introduction to XQuery". In *Proc. XSYM*, pages 5–20, 2004.

[24] R. Hull and J. Su. "On Accessing Object-oriented Databases: Expressive Power, Complexity, and Restrictions". In *Proceedings of the 1989 ACM SIGMOD International Conference on Management of Data (SIGMOD'89)*, pages 147–158, 1989.

[25] R. Hull and J. Su. "Algebraic and Calculus Query Languages for Recursively Typed Complex Objects". *J. Comput. Syst. Sci.*, **47**(1):121–156, 1993.

[26] G. Jaeschke and H.-J. Schek. "Remarks on the Algebra of Non First Normal Form Relations". In *Proc. PODS'82*, pages 124–138, 1982.

[27] D. S. Johnson. "A Catalog of Complexity Classes". In J. van Leeuwen, editor, *Handbook of Theoretical Computer Science*, volume 1, chapter 2, pages 67–161. Elsevier Science Publishers B.V., 1990.

[28] C. Koch, S. Scherzinger, N. Schweikardt, and B. Stegmaier. "Schema-based Scheduling of Event Processors and Buffer Minimization for Queries on Structured Data Streams". In *Proceedings of the 30th International Conference on Very Large Data Bases (VLDB)*, Toronto, Canada, 2004.

[29] G. Kuper and M. Y. Vardi. "A New Approach to Database Logic". In *Proceedings of the 3rd ACM SIGACT-SIGMOD-SIGART Symposium on Principles of Database Systems (PODS'84)*, pages 86–96, 1984.

[30] G. Kuper and M. Y. Vardi. "On the Complexity of Queries in the Logical Data Model". *Theor. Comput. Sci.*, **116**(1&2):33–57, 1993.

[31] L. Libkin. *Elements of Finite Model Theory*. Springer, 2004.

[32] L. Libkin and L. Wong. "Query Languages for Bags and Aggregate Functions". *Journal of Computer and System Sciences*, **55**(2):241–272, 1997.

[33] B. Ludäscher, P. Mukhopadhyay, and Y. Papakonstantinou. "A Transducer-Based XML Query Processor". In *Proceedings of the 28th International Conference on Very Large Data Bases (VLDB)*, pages 227–238, 2002.

[34] A. Marian and J. Siméon. "Projecting XML Documents". In *Proceedings of the 29th International Conference on Very Large Data Bases (VLDB)*, pages 213–224, 2003.

[35] C. H. Papadimitriou. *Computational Complexity*. Addison-Wesley, 1994.

[36] J. Paredaens and D. Van Gucht. "Possibilities and Limitations of Using Flat Operators in Nested Algebra Expressions". In *Proc. PODS*, pages 29–38, 1988.

[37] L. Segoufin. "Typing and Querying XML Documents: Some Complexity Bounds". In *Proceedings of the 22nd ACM SIGACT-SIGMOD-SIGART Symposium on Principles of Database Systems (PODS)*, pages 167–178, 2003.

[38] D. Suciu and V. Tannen. "A Query Language for NC". In *Proceedings of the ACM SIGACT-SIGMOD-SIGART Symposium on Principles of Database Systems (PODS'94)*, pages 167–178, 1994.

[39] V. Tannen, P. Buneman, and L. Wong. "Naturally Embedded Query Languages". In *Proc. of the 4th International Conference on Database Theory (ICDT)*, pages 140–154, 1992.

[40] M. Y. Vardi. "The Complexity of Relational Query Languages". In *Proc. 14th Annual ACM Symposium on Theory of Computing (STOC'82)*, pages 137–146, San Francisco, CA USA, May 1982.

[41] S. Vorobyov and A. Voronkov. "Complexity of Nonrecursive Logic Programs with Complex Values". In *Proceedings of the 17th ACM SIGACT-SIGMOD-SIGART Symposium on Principles of Database Systems (PODS'98)*, 1998.

[42] A. Voronkov. Personal communication, 2004.

[43] L. Wong. "Normal Forms and Conservative Extension Properties for Query Languages over Collection Types". *Journal of Computer and System Sciences*, **52**(3):495–505, 1996.

[44] World Wide Web Consortium. "XQuery 1.0 and XPath 2.0 Formal Semantics. W3C Working Draft (Aug. 16th 2002), 2002. http://www.w3.org/TR/query-algebra/.

[45] "XML Query Use Cases. W3C Working Draft 02 May 2003", 2003. http://www.w3.org/TR/xmlquery-use-cases/.

# APPENDIX

**Proof of Proposition 3.2 (Sketch).** For simplicity, we will here assume that all tuples are pairs, but the proof immediately generalizes to tuples of higher arity.

We assume complex values given as strings constructed using symbols from alphabet $\Sigma$ consisting of $\langle, \rangle, \{, \}, \text{``,''}$, and character symbols for atomic values.

For example, the value

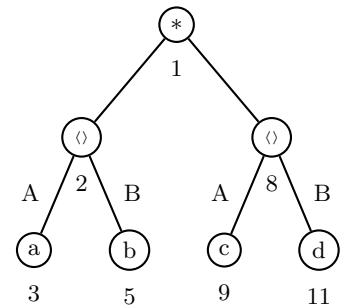

of type

$$\{\langle A : \text{Dom}, B : \text{Dom}\rangle\}$$

is represented as string "$\{\langle a,b\rangle, \langle c,d\rangle\}$".

Given a complex value $v$, we identify (set-, pair-, and atomic) terms of $v$ (i.e., nodes of the tree shown above) by the index of the first symbol of the term in the input string. For instance, the root node is identified with index 1 because the opening curly brace is the first symbol of the input string.

Let $I_v = \{1, \ldots, |v|\}$. Let *flat* be a function that maps every complex value $v$ of type $\tau$ to a relational structure $\langle Set, Pair, Atomic\rangle$ with relations $Set \subseteq I_v^2$, $Pair \subseteq I_v^3$, and $Atomic \subseteq I_v \times Dom$, where $\langle x,y\rangle \in Set$ iff there is a set-term $t_v(x)$ in $v$ that has a term $t_v(y)$ as member, $\langle x,y,z\rangle \in Pair$ iff there is a pair-term $t = \langle t_1, t_2\rangle$ in $v$ with $t = t_v(x)$, $t_1 = t_v(y)$, and $t_2 = t_v(z)$, and $\langle x,w\rangle \in Atomic$ iff there is an atomic term $t = w$ in $v$ with $t = t_v(x)$.

For the example value discussed above,

$$\begin{aligned}
Atomic &= \{\langle 3,a\rangle, \langle 5,b\rangle, \langle 9,c\rangle, \langle 11,d\rangle\} \\
Set &= \{\langle 1,2\rangle, \langle 1,8\rangle\} \\
Pair &= \{\langle 2,3,5\rangle, \langle 8,9,11\rangle\}
\end{aligned}$$

We have the power of $TC_0$ at hand to define a reduction from our input strings to the flat relations. We will not go into the details of a $TC_0$ reduction (cf. [4]), these are technical but in this case easy. The only point worth mentioning is that we can check whether two indexes $i, j$ are the left and right delimiters of a set or tuple. We show this in FO-order logic with majority quantifiers (FOM). By [4], $TC_0 =$ FOM. It is also known [4] that FOM can express predicates $x = y + z$ and $x = \#y\,\phi(y)$, such that $x$ is the number of positions $y$ for which $\phi(y)$ holds.

$$\begin{aligned}
\text{set-node}(i,j) \;:=\;\; & Q_{\{}(i) \wedge Q_{\}}(j) \wedge \\
& x = \#u\bigl(Q_{\{}(u) \wedge i < u < j\bigr) \wedge \\
& y = \#u\bigl(Q_{\}}(u) \wedge i < u < j\bigr) \wedge x = y \\
\text{tuple-node}(i,j) \;:=\;\; & Q_{\langle}(i) \wedge Q_{\rangle}(j) \wedge \\
& x = \#u\bigl(Q_{\langle}(u) \wedge i < u < j\bigr) \wedge \\
& y = \#u\bigl(Q_{\rangle}(u) \wedge i < u < j\bigr) \wedge x = y
\end{aligned}$$

where $(Q_a)_{a \in \Sigma}$ represents the input string and $Q_a(i)$ is true iff symbol $a$ is at position $i$ of the input string. (That is, these formulae state that $i, j$ are positions of symbols with matching opening and closing delimiters and the number of opening delimiters occurring between $i$ and $j$ is the same as the number of closing delimiters.)

Atomic nodes can already be defined in FO. Let "node" denote nodes of any of the three kinds. Now, for instance,

$$\phi_{Set}(i,j) := \exists i', j'\; \text{set-node}(i, i') \wedge \text{node}(j, j') \wedge i < j \wedge j' < i' \wedge$$
$$\neg \exists k, k'\; \text{node}(k, k') \wedge i < k < j \wedge j' < k' < i'.$$

This formula states that $i$ is the identifier of a set-node and $j$ the identifier of one of its children.

Let the $\mathcal{M}_\cup[\sigma]$ query $V_\tau$ for the type of the input data be defined inductively as

$$\begin{aligned}
V_{\text{Dom}} \;:=\;\; & Atomic \circ \text{map}(\langle 1 : \pi_1, 2 : \pi_2 \circ \text{sng}\rangle) \\
V_{\langle A:\tau_1, B:\tau_2 \rangle} \;:=\;\; & Pair \circ \text{map}(\langle 1 : \pi_1, 2 : V_{\tau_1}|\pi_2 \times V_{\tau_2}|\pi_3\rangle) \\
V_{\{\tau\}} \;:=\;\; & Set \circ \langle 1 : \text{map}(\pi_1), 2 : \text{id}\rangle \circ \text{pairwith}_1 \circ \\
& \text{map}\bigl(\langle 1 : \pi_1, 2 : \pi_2|\pi_1 \circ (\text{id} \times V_\tau) \circ \\
& \sigma_{1=2.1} \circ \text{map}(\pi_{2.2}) \circ \text{flatten} \circ \text{sng}\rangle\bigr)
\end{aligned}$$

where $S|v = \langle 1:v, 2:S\rangle \circ \text{pairwith}_S \circ \sigma_{1=2.1} \circ \text{map}(\pi_{2.2})$.

For our example, we get $V_\tau$ as shown in Figure 7.

It is not difficult to verify that for every complex value $v$ of type $\tau$, $V_\tau(flat(v)) = \{\langle 1:i, 2:\{v\}\rangle\}$, where $i$ is the identifier of the root of $v$, and that $V' := V_\tau \circ \text{map}(\pi_2) \circ \text{flatten}$ computes $\{v\}$.

By Theorem 2.5, for every $\mathcal{M}_\cup[\sigma]$ query from flat relations to flat relations there is an equivalent relational algebra query. Thus, for any Boolean $\mathcal{M}_\cup[\sigma]$ query $Q$, there is a relational algebra query $Q' \equiv V' \circ \text{map}(Q) \circ \text{flatten}$. Of course,

$$Q(v) \Leftrightarrow (V' \circ \text{map}(Q) \circ \text{flatten})(flat(v)) \Leftrightarrow Q'(flat(v)).$$

For a fixed query $Q$ (and thus a fixed type $\tau$), $Q'$ is fixed and can be evaluated on a (flat relational) database in $AC_0$ (cf. e.g. [31, 3]) and thus in $TC_0$. Preprocessing function $flat$ is in $TC_0$, so we can compose these two steps and get a $TC_0$ overall bound. □

$$
\begin{aligned}
V_{\text{Dom}} &= \{\langle 3, \{a\}\rangle, \langle 5, \{b\}\rangle, \langle 9, \{c\}\rangle, \langle 11, \{d\}\rangle\} \\
V_{\langle A:\text{Dom}, B:\text{Dom}\rangle} &= \{\langle 2, 3, 5\rangle, \langle 8, 9, 11\rangle\} \circ \text{map}(\langle 1 : \pi_1, 2 : V_{\text{Dom}}|\pi_2 \times V_{\text{Dom}}|\pi_3\rangle) \\
&= \{\langle 2, V_{\text{Dom}}|3 \times V_{\text{Dom}}|5\rangle, \langle 8, V_{\text{Dom}}|9 \times V_{\text{Dom}}|11\rangle\} \\
&= \{\langle 2, \{a\} \times \{b\}\rangle, \langle 8, \{c\} \times \{d\}\rangle\} \\
&= \{\langle 2, \{\langle a, b\rangle\}\rangle, \langle 8, \{\langle c, d\rangle\}\rangle\} \\
V_{\{\langle A:\text{Dom}, B:\text{Dom}\rangle\}} &= \{\langle 1, 2\rangle, \langle 1, 8\rangle\} \circ \langle 1 : \text{map}(\pi_1), 2 : \text{id}\rangle \circ \text{pairwith}_1 \circ \\
&\quad \text{map}\bigl(\langle 1 : \pi_1, 2 : \pi_2|\pi_1 \circ (\text{id} \times V_{\langle A:\text{Dom}, B:\text{Dom}\rangle}) \circ \sigma_{1=2.1} \circ \text{map}(\pi_{2.2}) \circ \text{flatten} \circ \text{sng}\rangle\bigr) \\
&= \langle 1 : \{1\}, 2 : \{\langle 1, 2\rangle, \langle 1, 8\rangle\}\rangle \circ \text{pairwith}_1 \circ \\
&\quad \text{map}\bigl(\langle 1 : \pi_1, 2 : \pi_2|\pi_1 \circ (\text{id} \times V_{\langle A:\text{Dom}, B:\text{Dom}\rangle}) \circ \sigma_{1=2.1} \circ \text{map}(\pi_{2.2}) \circ \text{flatten} \circ \text{sng}\rangle\bigr) \\
&= \{\langle 1 : 1, 2 : \{\langle 1, 2\rangle, \langle 1, 8\rangle\}\rangle\} \circ \\
&\quad \text{map}\bigl(\langle 1 : \pi_1, 2 : \pi_2|\pi_1 \circ (\text{id} \times V_{\langle A:\text{Dom}, B:\text{Dom}\rangle}) \circ \sigma_{1=2.1} \circ \text{map}(\pi_{2.2}) \circ \text{flatten} \circ \text{sng}\rangle\bigr) \\
&= \{\langle 1 : 1, 2 : \{2, 8\} \circ (\text{id} \times V_{\langle A:\text{Dom}, B:\text{Dom}\rangle}) \circ \sigma_{1=2.1} \circ \text{map}(\pi_{2.2}) \circ \text{flatten} \circ \text{sng}\rangle\} \\
&= \{\langle 1 : 1, 2 : (\{2, 8\} \times \{\langle 2, \{\langle a, b\rangle\}\rangle, \langle 8, \{\langle c, d\rangle\}\rangle\}) \circ \sigma_{1=2.1} \circ \text{map}(\pi_{2.2}) \circ \text{flatten} \circ \text{sng}\rangle\} \\
&= \{\langle 1 : 1, 2 : \{\langle 2, \langle 2, \{\langle a, b\rangle\}\rangle\rangle, \langle 8, \langle 8, \{\langle c, d\rangle\}\rangle\rangle\}\rangle \circ \text{map}(\pi_{2.2}) \circ \text{flatten} \circ \text{sng}\rangle\} \\
&= \{\langle 1 : 1, 2 : \{\{\langle a, b\rangle\}, \{\langle c, d\rangle\}\} \circ \text{flatten} \circ \text{sng}\rangle\} \\
&= \{\langle 1 : 1, 2 : \{\{\langle a, b\rangle, \langle c, d\rangle\}\}\rangle\}
\end{aligned}
$$

Figure 7: $V_\tau$ for the running example.